\documentclass[12pt]{iopart}
\usepackage{iopams}
\usepackage{siunitx}
\expandafter\let\csname equation*\endcsname\relax

\expandafter\let\csname endequation*\endcsname\relax

\usepackage{amsmath}
\usepackage[normalem]{ulem}

\usepackage{graphicx}
\usepackage{color}
\usepackage{dcolumn}
\usepackage{bm}
\usepackage{hyperref}
\begin{document}

\title[Correlation functions of non-Markovian systems
out of equilibrium]{Correlation functions of non-Markovian systems
out of equilibrium: Analytical expressions beyond single-exponential memory}
\author{Timo J. Doerries}
\ead{timo.doerries@uni-potsdam.de}
\address{Institut f\"ur Physik und Astronomie, Universit\"at Potsdam, Potsdam-Golm, Germany}
 \author{Sarah A.M. Loos}%
 \ead{sarah.loos@uni-leipzig.de}
  \address{Institut f\"ur Theoretische Physik, Universit\"at Leipzig, Leipzig, Germany}
\author{Sabine H.L. Klapp}
\ead{klapp@physik.tu-berlin.de}
\address{%
 Institut f\"ur Theoretische Physik, Technische Universit\"at Berlin, Berlin, Germany}%

\begin{abstract}
This paper is concerned with correlation functions of stochastic systems
with memory, a prominent example being a molecule or colloid moving 
through a complex (e.g., viscoelastic) fluid environment. Analytical investigations
of such systems based on non-Markovian stochastic equations are notoriously difficult.
A common approximation is that of a single-exponential memory, corresponding
to the introduction of one auxiliary variable coupled to the Markovian dynamics of the main variable.
As a generalization, we here investigate a class of ``toy'' models with altogether three degrees of freedom, giving rise
to more complex forms of memory. Specifically, we consider, mainly on an analytical basis,
the under- and overdamped motion of a colloidal particle coupled linearly
to two auxiliary variables, where the coupling between variables can be either reciprocal or non-reciprocal.
Projecting out the auxiliary variables, we obtain
non-Markovian Langevin equations with friction kernels and colored noise, whose structure is similar to that
of a generalized Langevin equation. For the present systems, however, 
the non-Markovian equations may violate the fluctuation-dissipation relation as well as detailed balance, indicating that the systems are out of equilibrium.
We then study systematically the connection between the coupling topology of the underlying Markovian system
and various autocorrelation functions.
We demonstrate that already two auxiliary variables
can generate surprisingly complex (e.g., non-monotonic or oscillatory) memory and correlation functions. Finally, we show that a minimal overdamped model with two auxiliary variables
and suitable non-reciprocal coupling yields correlation functions resembling those describing hydrodynamic
backflow in an optical trap.
\end{abstract}


%
%
%
%
%

\section{Introduction}
In recent years, the investigation of non-Markovian (``memory") effects in physical systems
has (re-)gained strong interest. ``Non-Markovianity" implies that the dynamics of one (or several) ``relevant" degree of freedom depends, to some extent, on its past. Such systems occur
in many areas of physics, such as the physics of strongly correlated soft matter (e.g., particles in viscous suspensions \cite{raikher2010theory,berner2018oscillating} or glass-forming liquids \cite{gotze2008complex}), in active \cite{nagai2015collective,PhysRevLett.121.078003,kursten2017giant,marchetti2013hydrodynamics,narinder2018memory}
and biological matter \cite{aguilar2018critical,mitterwallner2020non},
in reaction kinetics, protein dynamics \cite{lange2006generalized}, neurophysics, quantum optics \cite{carmele2019non}, but also in systems subject to feedback control with time delay \cite{loos2019heat,khadka2018active,khadem2019delayed,scholl2016control}.
Recent research on non-Markovian effects concerns, e.g., the complex behavior of dynamical correlation functions exhibiting multi-exponential \cite{mitterwallner2020non} and oscillatory \cite{berner2018oscillating} decay, 
molecular friction \cite{straube2020rapid},
unusal transport properties such as (anomalous) diffusion \cite{klages2008anomalous,hofling2013anomalous}, mobility \cite{gernert2016feedback} and oscillating particle currents \cite{lichtner2010feedback}, spontaneous modes of motion in active matter \cite{PhysRevLett.121.078003}, non-Markovian (quantum) phase transitions \cite{Kyaw_2020}, and thermodynamic properties of stochastic non-Markovian systems \cite{munakata2014entropy,loos2020thermodynamic,debiossac2020thermodynamics,di2020thermodynamic}. 

In theoretical approaches, non-Markovianity naturally arises as a result of a coarse-graining
procedure, when the (typically ``slow'') variable of interest is coupled to other,
less relevant (and typicaller ``faster'') ones. Unless the timescales of the slow and fast variables are completely separated,
the dynamics of the slow variable will involve, to some extent, a history dependency, i.e. non-Markovianity.
For classical systems, this is directly seen within
the celebrated Mori-Zwanzig approach
\cite{zwanzig1961memory,mori1965transport} yielding a generalized Langevin
equation (GLE) with a ``friction'' kernel and colored noise. In fact, both the kernel and the colored noise
induce a history dependency. Further, GLEs are used to model the impact of complex environments, such as ``particles" in the interior of a cell \cite{hofling2013anomalous,daldrop2017external} (as an alternative to other routes such as non-Gaussian diffusivity 
\cite{metzler2017gaussianity,wang2012brownian,chechkin2017brownian}
or fractional Brownian motion \cite{Yag58,jeonvivi,szymanski2009elucidating}).

Clearly, non-Markovian dynamics renders any analytical (and numerical, see, e.g., \cite{jung2018generalized}) 
treatment of a given model system more involved. For that reason, memory effects are often described on the basis of simple ansatzes, a popular one
being GLEs with friction kernels composed of exponentials. In the spirit of the Markovian embedding method \cite{siegle2010markovian}, the resulting GLEs can then be rewritten as a set of a (typically finite) number of Markovian equations involving $n$ auxiliary variables.
Many applications are based, in fact, on single-exponential kernels and thus, $n=1$ auxiliary variables (see, e.g., \cite{raikher2010theory}), but several recent studies consider $n=2$ (see, e.g., \cite{kappler2019non}). For systems in thermal equilibrium with a bath, the corresponding colored noise correlations are typically assumed
to be determined by the fluctuation dissipation relation (FDR)  \cite{zwanzig2001nonequilibrium} (sometimes referred to as fluctuation-dissipation theorem of second kind), stating that the noise correlations and the friction kernel are proportional to one another. However, there are also pure non-equilibrium models with one auxiliary variable, such as the active Ornstein-Uhlenbeck process 
\cite{Shankar2018,Caprini2019,Dabelow2019,Martin2020} describing the motion of an active particle
with fluctuating self-propulsion. 

Most investigations so far were focused on {\em specific} equilibrium or non-equilibrium systems. In contrast, the goal
of the present paper is to systematically explore, on a more general level, the autocorrelation functions of the position (PACF) and velocity (VACF) of classical stochastic systems with friction kernels and colored noise
generated by two auxiliary variables, i.e., $n=2$. The VACF and PACF are particularly interesting quantities as they are experimentally (or numerically) accessible in a large variety of soft matter systems.
For the sake of an analytical treatment, we focus on purely linear systems. Different from earlier studies, however, we do not assume that the couplings between the main variable and the auxiliary ones stem from an underlying Hamiltonian. This allows us to model both, equilibrium and non-equilibrium systems. In particular, we discuss both {\em reciprocal} coupling schemes (which, in the case of position variables, are typical of conservative forces fulfilling Newton's third law) and {\em non-reciprocal}
coupling schemes, which are one possible ingredient for modelling non-equilibrium systems \cite{loos2020thermodynamic,PhysRevE.101.022120}. 
In summary, we discuss a class of models whose (potential) non-equilibrium character is not induced by external forces or fields, but rather is of intrinsic character. Specifically, non-equilibrium in our models
arises through non-reciprocal coupling and/or different bath temperatures in the Markovian description, which is then reflected by a broken FDR \cite{zwanzig2001nonequilibrium} in the non-Markovian description.
%
%

We consider a wide range of coupling constants, including specific topologies of the network of (main and auxiliary) variables defined by setting some coupling constants to zero. For these, we outline ``maps" relating various behaviors of friction kernels in the non-Markovian picture and correlation functions. 
By this, we provide a guide how to capture specific features of measured correlation functions by a simple, analytically accessible model.
%
Indeed, the case $n=2$ introduces substantial additional complexity as compared to the single-exponential case, such as oscillatory or non-monotonic friction kernels and autocorrelation functions describing competing relaxational effects. For example, in case that the kernel has a maximum at a finite time, this can model a preferred ``delay" time \cite{loos2020thermodynamic}
(notice that the limit $n\to\infty$ corresponds to the case
of a delta-distributed, i.e., discrete delay time often used in time-delayed feedback control \cite{loos2019fokker}).
Aside of modelling different friction kernels, our approach allows for identifying the impact
of colored noise on the resulting correlation functions. 
As a concrete example, we show that a minimal overdamped model with two auxiliary variables
and suitable coupling yields correlation functions resembling those describing hydrodynamic
backflow in an optical trap \cite{franosch2011resonances}.

We close this introduction by a brief outline of the remainder of the paper and an overview of the main results. In Sec.~\ref{sec:model} we introduce our Markovian models composed of three linear (under- or overdamped) Langevin equations with $n=2$ auxiliary variables. 
While the idea of the paper is to propose a preferably general framework, which can then be applied to specific real systems (such as the one studied in Sec.~\ref{sec:appl}), we also introduce, for the sake of parameter reduction, specific coupling topologies.
Further, we relate our approach to models  studied in recent literature on viscoelastic, active and granular systems in Sec.~\ref{sec:related}. Section~\ref{sec:projection} is concerned with the corresponding non-Markovian equations for the physically relevant variables, that is, velocity or position.
We focus on the shape of the resulting friction kernel depending on the coupling topology. We show that, despite of the restriction to $n=2$, rather complex (e.g., oscillatory and non-monotonic) kernels can occur. Section~\ref{sec:correlations} is devoted to the calculation of measurable dynamical quantities, particularly the VACF and PACF. We provide full ``maps" illustrating a range of behaviors of 
time-dependent autocorrelation functions
of the velocity (underdamped case) or position (overdamped) case, including  oscillatory behavior. In Sec.~\ref{sec:fdr} we discuss our models in the light of two common measures of (non-)equilibrium, that is, the FDR and detailed balance. We show that, for most parameter settings, the described
systems are actually out of equilibrium, contrary to standard models based on generalized Langevin equations. Section~\ref{sec:appl} describes the application to a real colloidal system. This section demonstrates, at the same time, that a simpler model with $n=1$ would not be sufficient
to fit the experimental data. Our conclusions are summarized in Sec.~\ref{sec:outlook}. The paper includes several appendices comprising technical details.

\section{Markovian and non-Markovian models}
\subsection{\label{sec:model}Markovian models}
In this section we introduce sets of Langevin equations which are inspired by (yet not restricted to)
the physical situation of a colloid moving in one dimension 
in a solvent characterized by the friction constant $\gamma_0$ and
temperature $T_0$. The position and velocity of the colloid are given by $x$ and 
$v=\dot x$, respectively. Besides these physical variables, the equations 
of motion further contain couplings to auxiliary variables $y_i$, $i\in\{1,2,\ldots,n\}$.
Each auxiliary variable $y_i$ is exposed to its own thermal bath with temperature 
$T_i$. For the sake of a fully analytical treatment, we also assume that the coupling among all variables
is linear.

We here restrict ourselves to systems involving two auxiliary variables, $n=2$. Indeed, as we will show in the subsequent sections, already
this case allows to describe non-Markovian correlations with features markedly different from the ``standard" case of one 
auxiliary variable.
Focusing first on underdamped motion (with mass $m=1$), we specifically consider the equations
\begin{subequations}
	\label{eq:markovallNoMatrix}
\begin{align}
	\dot x~&=v\label{eq:markovall_x}\\
	\dot v~&=-\gamma_0v-\kappa x+k_1y_1+k_2y_2 + \xi_0\label{eq:markovall_v}\\
	\dot y_i&=-a_i y_i+b_iv+d_i y_j+\xi_i,\, i=1,2,\,j\neq i,\label{eq:markovall_yi}
\end{align}
\end{subequations}
where $\xi_\alpha$ (with $\alpha=0,1,2$) are zero-mean Gaussian white noises with correlations
\begin{align}
    \langle \xi_\alpha(t+\Delta t)\xi_\beta(t)\rangle= 2k_B T_\alpha \gamma_\alpha\delta_{\alpha\beta}\delta(\Delta t).
 \label{eq:shortC}
\end{align}
In Eq.~(\ref{eq:shortC}), the brackets $\langle \cdots \rangle$ denote an ensemble average over different noise realizations, $k_B$ is the Boltzmann constant, and $\delta_{\alpha \beta}$ is the Kronecker delta, which is
one for $\alpha=\beta$ and zero otherwise. Without loss of generality, we scale from now the constants $\gamma_1$ and $\gamma_2$ into the temperatures $T_{1,2}$ (such that in the following equations $T_{i}$ must be replaced by $\gamma_i T_{i}$ to return to usual temperatures).

We now discuss, on a quite general level, coupling properties of the network described by Eq.~(\ref{eq:markovallNoMatrix}). Indeed, our goal here is to provide a versatile framework which can then be specified (in terms of the coupling constants) to certain real systems, such as the optically trapped colloid in a viscous solvent discussed in Sec.~\ref{sec:appl}, and other types of systems studied earlier in the literature (see Sec.~\ref{sec:related}).

To start with, all coupling constants appearing in Eqs.~(\ref{eq:markovallNoMatrix}) are real. The case $\kappa=0$ leads to the absence of any conservative, $x$-dependent force, while $\kappa>0$ describes the impact of a confining harmonic potential. 
In Eqs.~(\ref{eq:markovall_yi}) we assume $a_i>0$, corresponding to a confinement of the $y_i$-dynamics. 
The coupling constants $k_i$, $b_i$, and $d_i$ ($i=1,2$)
connecting $v$ 
and the auxiliary variables $y_i$ can have arbitrary sign. Importantly, we do not impose any restrictions on these constants.
Thus, the coupling can be {\em reciprocal}, corresponding to $k_i=b_i$ ($i=1,2$), and $d_1=d_2$, or {\em non-reciprocal}, where these symmetries are not fulfilled.
We note that the {\em interpretation} of the (non)reciprocity strongly depends on whether one considers the auxiliary variables as physical degrees of freedom. 
For example, if we consider in the underdamped model (\ref{eq:markovallNoMatrix}) the $y_i$ as velocities (of two additional particles), then all coupling terms between $v$ and $y_i$ 
are friction-like (i.e., non-conservative) and
non-reciprocal coupling implies asymmetric frictional forces. 
As another example, if both $y_i$ are positions, then the choice $d_1\neq d_2$ would correspond to harmonic ``forces'' which, however, do not fulfill Newton's third law and are, in that sense, non-conservative. 
These issues are not relevant for the resulting correlation functions of the main 
variable (here $v$), which is why we leave the question of whether the $y_i$ have a physical meaning, at this point open. However, this question does become important in the context of equilibrium conditions, see Sec.~\ref{sec:fdr}.

\begin{figure}
    \centering
   \includegraphics[width=.23\textwidth]{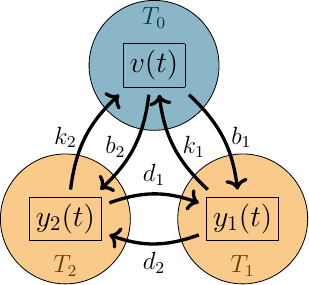}
    \caption{Schematic of the Markovian model system with $n=2$ auxiliary variables in the underdamped case.
    Black arrows indicate the (linear) coupling between variables.
    The velocity ($v$) of the colloid couples to two auxiliary variables $y_i$ ($i\in \{1,2\}$)
    with coupling strength $k_i$ and $b_i$, while the coupling between
    $y_i$ is determined by the constants $d_i$. Non-reciprocal coupling corresponds to $k_i\neq b_i$ and/or $d_1\neq d_2$. All variables are subject to thermal baths with temperatures $T_0$ and ${T}_i$. 
    }
    \label{fig:scheme}
\end{figure}

An illustration of the ``network" involving the three variables $v$, $y_1$ and $y_2$ [see Eqs.~(\ref{eq:markovallNoMatrix}) with $\kappa=0$] is given in Fig.~\ref{fig:scheme}.
As we will demonstrate, the coupling topology of the network, i.e., the set of coupling parameters, crucially determines the resulting non-Markovian dynamics and thus, the (measurable) correlation functions.
The most general case is the
``all-to-all" coupling topology (see Fig.~\ref{fig:scheme}), where all parameters $k_i$, $b_i$, and $d_i$, are non-zero. 
In addition, it is appropriate to introduce some special topologies (see insets in Fig.~\ref{fig:kernel}):
\begin{itemize}
\item ``Ring" topology, defined by $k_1=d_1=b_2=0$. This 
corresponds to a unidirectional coupling of the form $v\rightarrow y_1\rightarrow y_2\rightarrow v$ which is, by construction, non-reciprocal in all variables.
This topology naturally occurs when applying Markovian embedding to time-delayed feedback control \cite{loos2019fokker}.
\item ``Chain" topology, where $k_2=b_2=0$, that is, $v$ couples only to $y_1$. The latter then provides the connection towards the second auxiliary variable ($y_2$).
For this topology, one can still consider reciprocal and non-reciprocal cases depending on the remaining coupling constants. 
\item ``Center" topology, defined by $d_1=d_2=0$. Thus, there is no coupling {\em between} the auxiliary variables, while both of them
couple to $v$ (and vice versa) either reciprocally or non-reciprocally. 
\end{itemize} 
By considering these special topologies, we effectively reduce 
the number of parameters of the system. We will later demonstrate the resulting consequences for the dynamical (auto-)correlation functions, see Sec.~\ref{sec:correlations}.
We consider this topology-based analysis as a useful strategy in order to find a model system with least complexity (and parameters)
for given experimental data.

Finally, Eqs.~(\ref{eq:markovallNoMatrix}) can be easily rewritten to describe an overdamped system.
To this end, we replace in Eqs.~(\ref{eq:markovall_v}) and (\ref{eq:markovall_yi}) the variable $v$ (velocity) by $x$ (position)
and remove, at the same time, Eq.~(\ref{eq:markovall_x}). (Note that this is different from the usual overdamped {\em limit} of Eqs.~(\ref{eq:markovallNoMatrix}) 
because we wish the auxiliary variables in our overdamped model to be coupled to $x$ and not to $\dot x$).
Further, setting $\gamma_0+\kappa=a_0$
yields our Markovian model for overdamped motion,
\begin{subequations}
    \label{eq:markovover}
    \begin{align}
        \dot x&=-a_0x+k_1y_1+k_2y_2 + \xi_0\label{eq:markovover_x}\\
 	\dot y_i&=-a_i y_i+b_ix+d_i y_j+\xi_i,\,(i=1,2,\,j\neq i).\label{eq:markovover_yi}   
    \end{align}
\end{subequations}
Equation~(\ref{eq:markovover_x}) corresponds to the overdamped equation of motion for the colloidal position in presence of a harmonic potential
with stiffness $a_0$ and a friction constant of one. 
Regarding the entire network of three variables, a particularly simple interpretation
arises in the fully {\em reciprocal} case, that is, $k_i=b_i$, $d_1=d_2$. In this case, Eqs.~(\ref{eq:markovover_yi}) describe
a system of three colloidal particles with positions $x$, $y_1$ and $y_2$, which are mutually coupled by springs and (depending on the choice of $a_0$, $a_i$) are subject to additional harmonic traps. 
Thus, the coupling terms here relate to conservative forces.
In contrast, non-reciprocal coupling implies (as in the underdamped case with position-like auxiliary variables) the presence of non-conservative forces which break Newton's third law.

\subsection{\label{sec:related}Related models and fields of application}
We close this section by a brief comparison of our models with some related (low-dimensional) models studied in the literature, targeting different fields of application.
Starting from the underdamped case Eq.~(\ref{eq:markovallNoMatrix}) and setting the second auxiliary variable to zero ($y_2=0$), we essentially arrive at the ``Jeffrey model" of retarded friction, which has been
introduced in the context of (macroscopic) rheology. More recently, Raikher and Rusakov \cite{raikher2010theory} have used the resulting equations
to fit data for a microrheological experiment, where a micron-sized colloid ``trapped" in a harmonic potential (mimicking an optical trap) is moving through
a viscoelastic solution. It is well established that the single auxiliary variable leads to a (single-)exponential memory, where
the constant $a_1$ plays the role of the inverse Maxwell (stress) relaxation time. Our model extends the ansatz in \cite{raikher2010theory} by introducing a second auxiliary variable which may be coupled not only to $v$,
 but also to the first one. Further, underdamped models with two or three auxiliary variables have been used by Kowalik {\em et al.} \cite{kowalik2019memory} for molecules moving in viscoelastic fluids. In these models, the auxiliary variables are typically
 not exposed to a heat bath. The most significant differences to our systems, however, is that the couplings in \cite{kowalik2019memory} are assumed to be reciprocal, as it is typical of systems derivable from a Hamiltonian.
 Further, an application of the overdamped case involving one auxiliary variable (i.e., $y_2=0$) is the ``active Ornstein-Uhlenbeck model" \cite{Shankar2018,Caprini2019,Dabelow2019} 
 used in the field of active matter. In that model, $x$ represents the
 particle position and $y_1$ represents a fluctuating ``self-propulsion" velocity. The coupling is unidirectional, with 
 $k_1>0$, $b_1=0$, and $a_0>0$, $a_1>0$. For a more detailed discussion of models of this type, see \cite{loos2020thermodynamic,Martin2020}.
 Finally, we would like to mention a number of studies in the field of driven granular matter where models similar to ours with few degrees of freedom, including the case of altogether three variables consistent with our choice $n=2$, have been used.
 This includes, in particular, an ``itinerant oscillator model" for the angular velocity and position of a rotating tracer in dense, vibrated granular matter \cite{doi:10.1063/1.4928456} exhibiting anormalous diffusion \cite{PhysRevLett.114.198001}, and variants of that model
 \cite{baldovin2019langevin,PhysRevE.102.012908}.
  
%
%
%
\subsection{\label{sec:projection}Non-Markovian representation}
In parallel to the sets of (Markovian) Langevin equations given in Eqs.~(\ref{eq:markovallNoMatrix}) and (\ref{eq:markovover}), we consider the non-Markovian
equations for the physically relevant variables $v$ or $x$ that result from ``integrating out" (projecting) the auxiliary variables $y_i$, Eqs.~(\ref{eq:markovall_yi}) and (\ref{eq:markovover_yi}). To this end we use Laplace transformation techniques
as outlined in \ref{sec:derivationProjection}. We first consider the underdamped case. Starting from Eqs.~(\ref{eq:markovallNoMatrix}), one obtains 

\begin{align}
	\dot v &= -\int_{0}^{t}\Gamma (t-t')v(t')dt'-\kappa x+\xi_0+\xi_c
        \label{eq:solutionxAllCoupled}
\end{align}
involving the friction kernel $\Gamma(t)=2\gamma_0\delta(t)+\gamma(t)$ and the colored noise, $\xi_c=\xi_c(t)$ (see Eq.~ (\ref{<+label+>}) for an explicit expression). The non-trivial contribution to the frictional kernel is given by
\begin{align}
\gamma(t)&=\gamma^{+}(t)+\gamma^{-}(t),
\label{eq:gammaplusminus}
\end{align}
where
\begin{align}
	\gamma^{\pm}(t)&=g^{\pm}\exp\left[-(a_1+a_2\pm\omega)\frac{t}{2}\right]
	\label{eq:frictionKernelUnder}
\end{align}
with
\begin{align}
	g^{\pm}&=\mp\frac{(k_1b_1-k_2b_2)(a_1-a_2)}{2\omega}-\frac{k_1b_1+k_2b_2}{2}
  \pm\frac{k_1d_1b_2+k_2d_2b_1}{\omega}
	\label{eq:gplusminus}
\end{align}
and 
\begin{align}
	\omega&=\sqrt{(a_1-a_2)^2+4d_1d_2}~.
\label{eq:omega}
\end{align}
\subsubsection{Characteristic features of friction kernel and colored noise}
It is worth to discuss some general features of the function $\gamma(t)$ independent of a specific topology.
First, the function $\gamma(t)$ trivially vanishes when $k_i=0$ or $b_i=0$ ($i=1,2$), that is, when $v$ decouples from the auxiliary variables.
Second, $\gamma(t)$ does not depend on the noise terms (and thus, the temperatures $T_i$) of the auxiliary variables. This is why we can discuss the friction kernel separately from noise effects. The same holds for the constant $\kappa$ in Eq.~(\ref{eq:markovall_v}), that is, the kernel is independent of the strength of confinement (if present).
Third, the fact that $\gamma(t)$ involves two exponential functions (resulting from the presence of two auxiliary variables) allows for the possibility of a {\em non-monotonic} friction kernel with a minimum or maximum (at $t>0$), if the prefactors have different signs.
Clearly, such a non-monotonicity already represents a marked difference to the case of one auxiliary variable (i.e., single-exponential memory). 
Fourth, there are parameter ranges where $\gamma(t)$ contains {\em oscillatory} contributions. This occurs when $\omega$ defined in Eq.~(\ref{eq:omega}) becomes (purely) imaginary, that is,
\begin{align}
	\left(a_1-a_2\right)^2&<-4d_1d_2~.
\label{eq:gammaoscillations}
\end{align}
Since $a_1$ and $a_2$ are real, the left hand side is always positive (or zero).
Thus, oscillations can not occur if either $d_1$ or $d_2$ is zero, that is, if there is a {\em unidirectional} coupling between
the auxiliary variables, or if there is no coupling between $y_i$ at all. Further, for oscillations to occur, $d_1$ and $d_2$ have to have different signs, i.e. the coupling between the auxiliary variables has to be {\em non-reciprocal}.
The fifth general statement regarding $\gamma(t)$ concerns its limit at time zero. Here, one finds from Eq.~(\ref{eq:frictionKernelUnder})
$\lim_{t\to 0}\gamma(t)=-k_1b_1-k_2b_2$; thus, the function $\gamma(t)$ can have an ``offset". The latter solely depends on the coupling between $v$ and the auxiliary variables, and is strictly negative for reciprocal cases.

\begin{figure}
\centering
   \includegraphics[width=.7\textwidth]{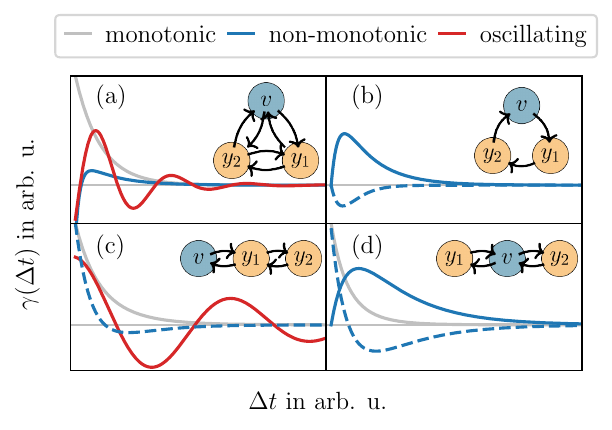}
    \caption{Examples of the function $\gamma(t)$ representing the non-trivial friction kernel in the underdamped case, see Eqs.~(\ref{eq:solutionxAllCoupled}) and (\ref{eq:gammaplusminus}). 
  The possible shapes of this function depend on the coupling topology. (a) ``all-to-all" topology, (b) ``ring" topology,
  (c) ``chain" topology, (d) ``center" topology. In each part, grey lines refer to monotonically decaying kernels, blue lines to non-monotonic kernels
  with one maximum (solid) or minimum (dashed) at a finite time, and red lines to oscillatory kernels. The corresponding parameters are given in Table~\ref{tab:kernel} in \ref{sec:kernelParameters}.}
    \label{fig:kernel}
 \end{figure}

Examples of the function $\gamma(t)$ are shown in Fig.~\ref{fig:kernel}. 
The most general case is the ``all-to-all" topology (Fig.~\ref{fig:kernel}(a)) for which one may observe monotonic decay (via a sum of exponentials), oscillatory behavior,
 or non-mononotic behavior characterized by one minimum or maximum at a finite time.
 The other panels in Fig.~\ref{fig:kernel} relate to the specific topologies introduced in Sec.~\ref{sec:model}.
Their main features can be summarized as follows:
\begin{itemize}
\item For the ``ring" topology (Fig.~\ref{fig:kernel}(b)): $\gamma(t)$ is {\em always} non-monotonic but never oscillatory (because of $d_1=0$). This is directly seen from the explicit expression
\begin{align}
	\gamma_{\text{ring}}(t)&=-\frac{2k_2b_1d_2}{a_1-a_2}\exp\left[-\frac{(a_1+a_2)t}{2}\right]\sinh\left(\frac{(a_1-a_2)t}{2}\right),
	\label{eq:gammaring}
\end{align}
revealing that $\gamma(t)$ starts from zero at $t=0$ and then approaches a maximum or minimum at a finite time. %
We note that the nonmonotonic behavior remains upon introducing more auxiliary variables (i.e., $n>2$) in the ring topology \cite{loos2020thermodynamic}.
\item For the ``chain" topology (Fig.~\ref{fig:kernel}(c)), one may observe non-monotonicity, oscillations, and (by default) an offset
at $t\rightarrow 0$. The explicit expression is given by
\begin{align}
	\gamma_{\text{chain}}(t)&=k_1b_1\exp\left[-\frac{(a_1+a_2)t}{2}\right]
	& \times \left(\frac{a_1-a_2}{\omega}\sinh\left(\frac{\omega t}{2}\right)-\cosh\left(\frac{\omega t}{2}\right)\right).
	\label{eq:gammachain}
\end{align}
\item In the ``center" topology (Fig.~\ref{fig:kernel}(d)), oscillations are not possible because of $d_i=0$. Still, non-monotonic behavior can occur, as seen from 
\begin{align}
	\gamma_{\text{center}}(t)&=-\sum_{i\in\{1,2\}}k_ib_i\exp\left[-a_it\right].
	\label{eq:gammastar}
\end{align}
\end{itemize}

Turning back to the non-Markovian Eq.~(\ref{eq:solutionxAllCoupled}), we now focus on the colored noise, $\xi_c$, whose explicit expression 
is given in Eq.~(\ref{eq:coloredNoise}). The colored noise involves a time integral, and thus, a history dependency. In this sense,
$\xi_c$ may be seen as an additional source of memory.
The corresponding noise correlation function in the stationary limit (for a derivation, see \ref{sec:derivationNoisecorr}), is given by 

\begin{align}
\lim_{t\to \infty}\langle \xi_c(t+\Delta t)\xi_c(t)\rangle &=\Xi^{+}(\Delta t)+\Xi^{-}(\Delta t)
\label{eq:noiseCorrelation}
\end{align}
with
	\begin{align}
\Xi^{\pm}(\Delta t)&=
2k_B\left(\frac{T_1c_1^\pm c_1^\pm}{a_1+a_2\pm\omega}+\frac{T_2c_2^{\pm} c_2^{\pm}}{a_1+a_2\pm \omega}
+\frac{T_1c_1^- c_1^+}{a_1+a_2}\right.\nonumber\\
&\hspace{4cm}
		\left.+\frac{T_2c_2^- c_2^+}{a_1+a_2}\right)\exp\left[-(a_1+a_2\pm\omega)\frac{\Delta t}{2}\right].
		\label{eq:noisecorrplusminus}
	\end{align}
In Eq.~(\ref{eq:noisecorrplusminus}), $c_i^{\pm}$ ($i=1,2$) are constants which are defined in Eq.~(\ref{eq:c12pm}).
The appearance of the temperatures $T_i$ in Eq.~(\ref{eq:noisecorrplusminus}) directly reflects that the colored noise and its correlations
only appear when the auxiliary variables are coupled to heat baths. We also note that
the noise correlations can be oscillatory in time. This occurs if $\omega$ defined in Eq.~(\ref{eq:omega}) becomes (purely) imaginary [see condition~(\ref{eq:gammaoscillations})] which, at the same
time, implies that the (non-trivial) friction kernel, $\gamma(t)$, oscillates. A more detailed discussion of relations between friction kernel
and noise correlations will be given in Sec.~\ref{sec:fdr}.

%

\subsubsection{Memory functions in the overdamped case}
Finally, we briefly consider the overdamped case. Following the same steps of calculation as in the underdamped case, we obtain from Eqs.~(\ref{eq:markovover})
\begin{align}
	\dot x &=-\int_0^t G(t-t')x(t')dt'+\xi_c(t)+\xi_0(t)~,
	\label{eq:xProjectedG}
\end{align}
where the kernel $G(t-t')$ is, by construction, identical to the function $\Gamma(t)$ appearing in the underdamped case, see Eq.~(\ref{eq:solutionxAllCoupled}). Physically, the convolution integral in Eq.~(\ref{eq:xProjectedG})
describes a retarded harmonic force. When modelling colloidal motion, however, one is rather interested in the ``friction kernel" where
the convolution involves not $x$, but $\dot x$. Starting from Eq.~(\ref{eq:xProjectedG}), this can be achieved through an integration by parts (see \ref{sec:derivationKernelover}). Focusing on the long time limit one obtains
\begin{align}
	\int_0^t\tilde{\Gamma}(t-t')\dot x(t')=
	-a_0'x+\xi_0+\xi_c,
	\label{eq:GLEoverdamped}
\end{align}
where $\tilde{\Gamma}(t)=2\delta(t)+\gamma_\text{o}(t)$, $\gamma_{\text{o}}(t)=\gamma_\text{o}^{+}(t)+\gamma_\text{o}^{-}(t)$,
and
\begin{align}
	\gamma_\text{o}^{\pm}(t)&=-2\frac{\gamma^{\pm}(t)}{a_1+a_2\pm w}
	\label{eq:frictionOverdamped}
\end{align}
with $\gamma^\pm(t)$ being defined in Eq.~(\ref{eq:frictionKernelUnder}). Due to direct connection to the friction kernel in the underdamped
model, many conclusions which we draw there also apply to the overdamped system. The main difference is the different sign.
 Further in Eq.~(\ref{eq:GLEoverdamped}), $a_0'=a_0-\gamma^+_\text{o}(0)-\gamma^-_\text{o}$,
and the colored noise, $\xi_c(t)$, is the same as that appearing in the underdamped case, see Eq.~(\ref{eq:coloredNoise}). 
\section{\label{sec:correlations}Correlation functions}
While the memory kernels discussed in Sec.~\ref{sec:projection} determine the non-Markovian equations of motion
(\ref{eq:solutionxAllCoupled}) and (\ref{eq:xProjectedG})
from a mathematical point of view, the quantities of prime {\em physical} interest (in terms of measurements) are rather
the (auto-)correlation functions of the system. Here we discuss, in particular,
the VACF and mean squared displacement (MSD) in the underdamped case, and briefly comment on the overdamped case.
%
\subsection{\label{sec:vacf}Closed expressions}
We start by considering the VACF in the long-time limit, defined by
\begin{align}
  c_V(\Delta t)&=\lim_{t\to \infty}\left\langle v(t)v(t+\Delta t)\right\rangle.
  \label{eq:vacf}
\end{align}

Our analytical calculation of $c_V$ relies on a Fourier transform of Eq.~(\ref{eq:solutionxAllCoupled}) and the calculation of the Green's function
	$\lambda(t)$ 
in the time domain to obtain the back transform. This is outlined in \ref{sec:green}. One obtains
\begin{align}
  c_V(\Delta t)=&\underbrace{\int_{-\infty}^t ds~ \lambda(s)\lambda(s+\Delta t) 2\gamma_0k_BT_0}_{:=c_V^\text{I}(\Delta t)} \nonumber \\
  &+\underbrace{\int_{-\infty}^t ds\int_{-\infty}^{t+\Delta t} ds'\lambda(s)\lambda(s')
	  \langle\xi_c(t-s) \xi_c(t+\Delta t -s')\rangle
}_{:=c_V^\text{II}(\Delta t)}.
  \label{eq:vacfInt}
\end{align}
On the right side of Eq.~(\ref{eq:vacfInt}), we defined the first function $c_V^\text{I}(\Delta t)$ which is proportional to $T_0$ (and independent of $\xi_c$). Inserting 
the Green's function Eq.~(\ref{eq:lambdareal}) and performing the integral we obtain (for $t\rightarrow\infty$)
\begin{align}
  c_V^\text{I}(\Delta t)=2\gamma_0k_BT_0\sum_{k=1,1\leq l \leq k}^{3}\frac{\epsilon_k\epsilon_l}{F_k+F_l}\left(e^{-F_k \Delta t}+e^{-F_l \Delta t}\right)~.
\label{eq:vcfA}
\end{align}
where the coefficients $\epsilon_k$ are given in Eq.~(\ref{eq:calcd}) and $F_k$ denote the poles of the Fourier transform of the Green's function $\hat\lambda$, see \ref{sec:green}. 

The second function defined in Eq.~(\ref{eq:vacfInt}), $c_V^\text{II}(\Delta t)$, involves the correlations of the colored noise [see Eq.~(\ref{eq:noiseCorrelation})]
and is thus linear in $T_1$ and $T_2$. The remaining integral is quite cumbersome, it can be performed
using a computer algebra program ({\em Mathematica 10.1)}\footnote{Script available from the authors upon request} yielding closed-form solutions.

Having obtained the VACF, the MSD can be calculated from the relation
\cite{hansen1990theory}
\begin{align}
\left\langle\left(\Delta x(t)\right)^2\right\rangle & \equiv
	\left\langle [x(t)-x(0)]^2\right\rangle
	= 2t\int_0^t dt'~\left(1-\frac{t'}{t}\right)c_V(t')~.
\label{eq:calcmsd}
\end{align}

The calculation becomes most convenient in the case that the auxiliary variables are {\em not} coupled to a heat bath, that is,
$T_1=T_2=0$, corresponding to $\xi_c(t)=0$, and thus, $c_V(t)=c_V^\text{I}(t)$ [see Eq.~(\ref{eq:vacfInt})]. In this case we obtain
from Eqs.~(\ref{eq:vcfA}) and (\ref{eq:calcmsd}), 
%
\begin{align}
\left\langle\left(\Delta x(t)\right)^2\right\rangle		
	&=4\gamma_0k_BT_0\sum_{k=1,1\leq l \leq k}^3\frac{\epsilon_k\epsilon_l}{F_k+F_l}\
	\left(\frac{e^{-F_kt}-1+F_k t}{F_k^2}
	\right.
		\left.+\frac{e^{-F_lt}-1+F_lt}{F_l^2}\right)~.
	\label{eq:msdUnder}
\end{align}
From Eq.~(\ref{eq:msdUnder}), we can readily see that the MSD grows linear in the
long time limit, corresponding to normal diffusion, i.e. $\lim_{t\to\infty}\langle (\Delta x(t))^2\rangle\propto t$.

Finally, for the overdamped system, the expression for the PACF follows directly from the expression for $c_V(t)$ by just changing the dynamic variable from $v$ to $x$.
This is since the projected equations (\ref{eq:solutionxAllCoupled}) for $\dot y$ and (\ref{eq:xProjectedG})
for $\dot x$, respectively, are formally equivalent.
%
\subsection{\label{sec:vacf_data} Discussion of autocorrelation functions}
We now discuss results for the correlation functions of an underdamped system.
Clearly, one is faced with a rather large parameter space spanned by the coupling parameters between the network variables, the constants $\gamma_0$ and $a_i$,
and the temperatures $T_\alpha$ ($\alpha=0,1,2$). After scanning broad ranges of these parameters, we here focus on a few, particularly interesting situations.
We first consider systems without colored noise, where the history dependency arises only through the friction kernels.
Indeed, we will see that this type of memory dominates the behavior of the correlation functions as compared to the effect of colored noise studied subsequently.

\subsubsection{Systems with white noise}
We start by considering a system with ``center" topology and $T_1=T_2=0$. The VACF is then given by $c_V(t)=c_V^\text{I}(t)$ [see Eqs.~(\ref{eq:vacfInt}) and (\ref{eq:vcfA})].
We recall that the center topology implies a (non-trivial) friction kernel $\gamma(t)$, which is
either monotonic or non-monotonic with one maximum or minimum at a finite time, see Eq.~(\ref{eq:gammastar}). The latter equation also indicates that {\em all} dynamical functions (memory kernel, VACF, MSD)
only depend on the products $k_ib_i$ rather than on each coupling constant individually [this is explicitly seen in Eq.~(\ref{eq:gammastar})]. 
For that reason, we can discuss the results in the plane spanned by $k_1b_1$ and $k_2b_2$ (for fixed $\gamma_0$, $a_i$). For each of these products 
we consider a range of positive and negative values. Notice that $k_ib_i<0$ corresponds to different signs of $k_i$ and $b_i$, which necessarily implies {\em 
non-reciprocal} coupling.

\begin{figure}
\begin{center}
  \includegraphics[width=\textwidth]{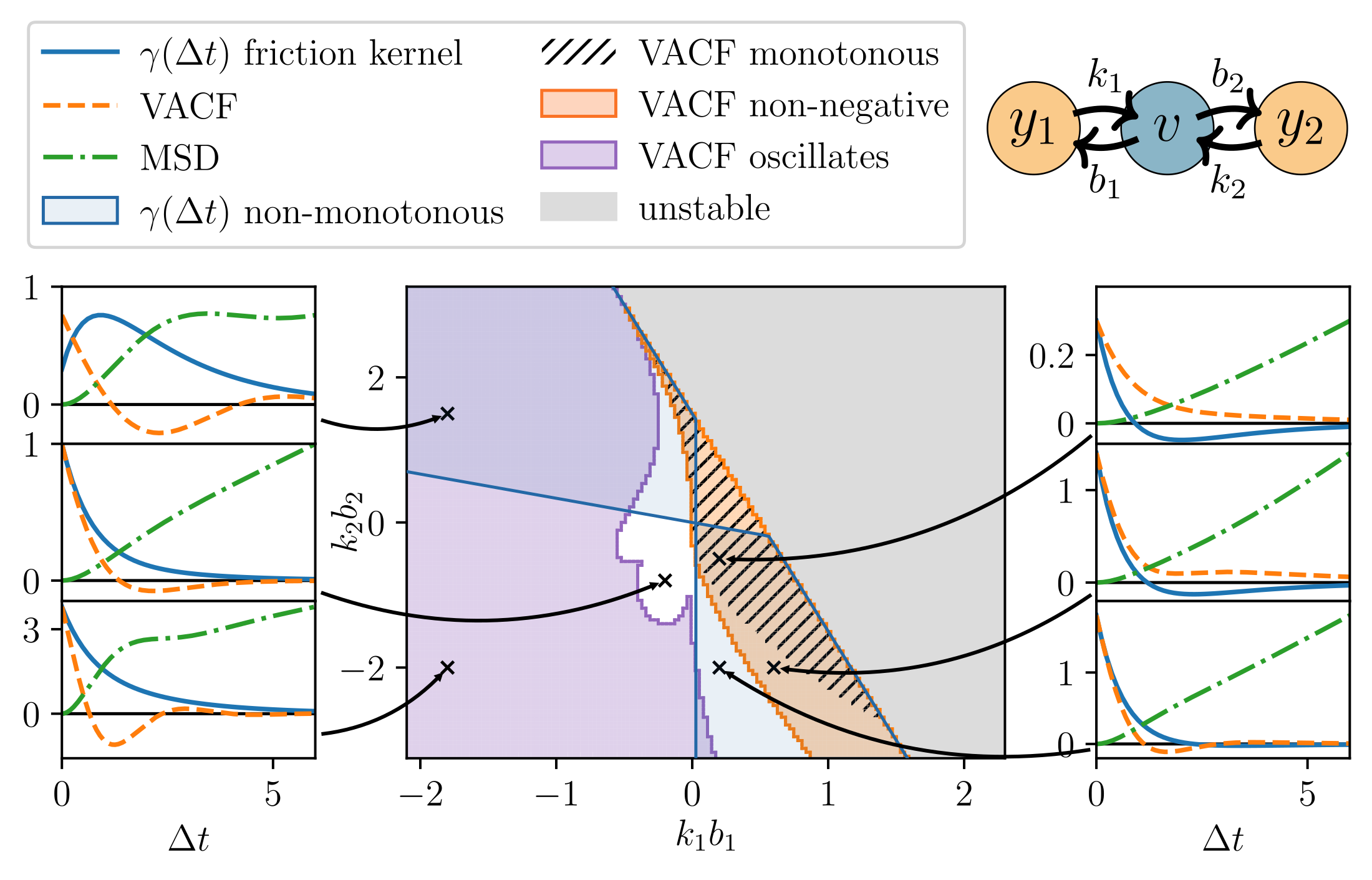}
    \caption{Overview of the various behaviors of the VACF and the MSD for underdamped systems with center topology in the absence of colored noise ($T_1=T_2=0$).
    The remaining constants are set to $T_0=0.5$, $\gamma_0=1$, $a_1=1/2$ and $a_2=3/2$. Also shown
    are the corresponding friction kernels. The correlation functions related to several examplary parameter combinations (indicated by crosses)
    are shown in the left and right panel.
    To enhance overall visibility, all VACFs and MSDs are multiplied by scaling factors given in 
    Table~\ref{tab:mapScales}. The various colored areas are explained in the main text.}
    \label{fig:VACFmap}
    \end{center}
 \end{figure}

In Fig.~\ref{fig:VACFmap} we present a ``map" illustrating the various behaviors of the VACF, MSD and friction kernel in the parameter plane spanned by $k_1b_1$ and $k_2b_2$. 
We can identify several regions.
In the grey region, the system is linearly unstable in the sense that the coefficient matrix of the deterministic part of the Markovian equations (\ref{eq:markovallNoMatrix}) for $v$ and $y_i$ 
has (at least) one eigenvalue, whose real part is positive. Outside the grey region, all eigenvalues have negative real parts. In the presence of (white) noise
acting on $v$, the variance of $v$, that is, the correlation function $\langle v(t)v(t)\rangle$ in its stationary limit, diverges when one
approaches the stability border (of the deterministic system) from the left. This, again, signals
the instability.

The remaining part of the plane can be divided according to different criteria. Focusing first on the VACF we find that
the simplest behavior occurs within the striped region, 
where the VACF decays monotonically from its initial value at $\Delta t=0$ to its long-time limit,
$\lim_{\Delta t\to\infty}c_V(\Delta t)=0$. Outside the striped, but still within the orange region, the VACF remains to be non-negative everywhere, but exhibits a minimum and a local
maximum at finite $\Delta t>0$, before eventually decaying to zero (see middle right panel for an example).
Within the violet region, the VACF shows
damped oscillatory behavior characterized by, at least, three roots.
We note that throughout this range, one of the products $k_ib_i$ is negative, implying that the coupling between the variables in the underlying Markovian network is non-reciprocal. 
Finally, outside the violet and orange regions, the VACF has one minimum at negative values, 
but does not oscillate in the sense that it has less than three roots.
%

Traditionally, minima in the VACF are often discussed in the context of strongly correlated liquids, where each particle (or molecule) is surrounded
by several neighbours (see, e.g., \cite{hansen1990theory}). These neighbours form a ``cage" around the particle considered, leading to backscattering and thus, negative values of the VACF at intermediate times. 
In the present, single-particle system, the minima rather result from the non-Markovian friction induced by the auxiliary variables.
To better understand the occurrence and location of the minima we note that, for the case of white noise considered here, there is an {\em exact} relationship between the time derivative
of the VACF and the function $\gamma(t)$. This relation, well known from the theory of liquids \cite{hansen1990theory,lange2006collective}, is derived by multiplying Eq.~(\ref{eq:solutionxAllCoupled}) by $v(t=0)$ and taking the average on both sides. 
Using that $\langle v(0)\xi_0\rangle=0$ and $\xi_c=0$, we obtain
\begin{align}
	\dot{c}_V(t) &= -\int_{0}^{t}\Gamma (t-t')c_V(t')dt'=-\gamma_0 c_V(t)-\int_{0}^{t}\gamma (t-t')c_V(t')dt',
        \label{eq:memoryequation}
\end{align}
where $\dot{c}_V(t)=\langle \dot{v}(t)v(0)\rangle$ and we have implicitly assumed that the system at $t=0$ is already in the steady state. 
According to Eq.~(\ref{eq:memoryequation}), a minimum of the VACF (i.e., $\dot{c}_V(t) =0$ at a finite time $t_{\text{min}}$) requires that the function
$\gamma(t)$ is non-local in time (which is trivially fulfilled in all cases considered here). 
%
Let us focus here on cases where $\gamma(t)$ has a {\em maximum} at a finite time and is entirely positive. 
This case is particularly interesting, e.g., in the context of time-delayed feedback control \cite{loos2020thermodynamic}; the maximum in $\gamma(t)$ then corresponds to a preferred ``delay" time.
As a major trend we observe an increase of $t_{\text{min}}$ with the time related to the maximum of the kernel (not shown here).
This may be understood such that the larger the delay, the later the VACF ``feels" the effect of the memory and changes its decay. 

A further interesting relationship emerges when we differentiate Eq.~(\ref{eq:memoryequation}) 
with respect to time and set $t=0$. This yields
\begin{align}
\frac{\ddot{c}_V(0)}{c_V(0)}=-\gamma_0\frac{\dot{c}_V(0)}{c_V(0)}-\gamma(0).
        \label{eq:vacf_curvature}
\end{align}
We here focus on the case $\gamma_0=0$, for which Eq.~(\ref{eq:vacf_curvature}) implies that the value $\gamma(0)$ alone controls the initial curvature of the VACF (note that $c_V(0)>0$). In particular, for positive $\gamma(0)$ 
the curvature is negative, and it becomes more and more negative with increasing $\gamma(0)$. This suggests, as a trend, that the change of sign of the VACF and the minimum $t_{\text{min}}$ occur the earlier, the larger
$\gamma(0)$ is. Physically speaking, large (positive) values
of $\gamma(0)$ lead to strong ``braking" of the motion, yielding a quick change of sign of the VACF. 
\begin{figure}
\begin{center}
	\includegraphics[scale=.8]{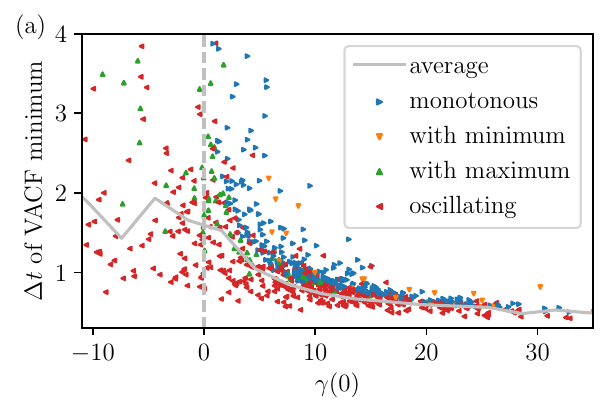}
	\includegraphics[scale=.8]{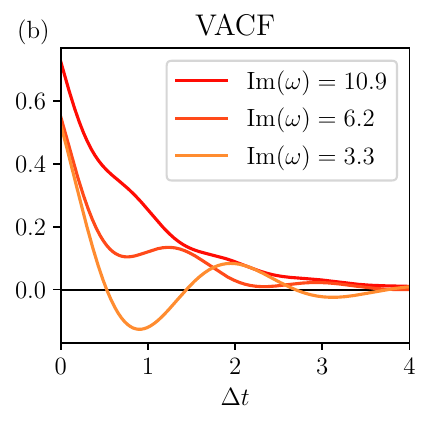}  
    \caption{
	    (a) Position of the first minimum of the VACF, $t_{\text{min}}$, as function of the initial value of the friction kernel. The data pertain 
    to $\gamma_0=0$ and $T_0=1$. Further, the constants $a_1$ and $a_2$ are randomly chosen from an uniform distribution
    between $0\dots5$, and all coupling parameters are randomly chosen from an
    uniform distribution from  $-5\dots 5$. 
    The different colours of data points correspond to different functional shapes
    of $\gamma(t)$. 
    (b) Examplary VACFs of underdamped systems with oscillatory friction kernels characterized by different oscillation frequencies (chain topology, $T_1=T_2=0$).
The parameters are $a_0 = 1$, $a_1 = 0.5$, $a_2 = 1.5$, $k_1 = -5$, $b_1 = 2$, $k_2=b_2=0$, $d_2 = 1$, and different values of $d_1$.}
    \label{fig:extrema_oscillations}
    \end{center}
 \end{figure}
That this is indeed a major trend can be seen in Fig.~\ref{fig:extrema_oscillations} where we plot the location of the first minimum as function of $\gamma(0)$ for a large set of coupling parameters.
Importantly, the corresponding network topologies include cases beyond the center topology (considered in Fig.~\ref{fig:VACFmap}), thereby enabling additional, particularly oscillating, shapes of $\gamma(t)$. Indeed, oscillatory kernels are absent for the center topology but emerge in the chain (or fully coupled) topology, see Eq.~(\ref{eq:gammachain}).
Despite the large variety of coupling parameters considered in Fig.~\ref{fig:extrema_oscillations}, we see that the general trend described above remains valid. It is most pronounced for monotonous kernels, but oscillating kernels yield similar behavior.
However, the observed trend breaks down for negative values of $\gamma(0)$, where the initial curvature of the
VACF is positive, and the direct connection to the minimum is lost.

A further question in this context is whether non-centered coupling topologies also yield other behaviors of the VACF not seen in Fig.~\ref{fig:VACFmap}. 
This question concerns, in particular, oscillatory kernels emerging for the chain topology, see Eq.~(\ref{eq:gammachain}), or for all-to-all coupling.
Figure~\ref{fig:extrema_oscillations} shows three examples of corresponding VACFs with different oscillation frequencies of $\gamma(t)$ i.e., different imaginary parts of the quantity $\omega$ defined in Eq.~(\ref{eq:omega}). In all cases, $T_1=T_2=0$. 
Several observations can be made. For small frequencies $\text{Im}(\omega)$, the VACF displays oscillatory behavior, similar to the cases discussed for the center topology (where, by definition, $\text{Im}(\omega)=0$). On the other hand, for high frequencies, the VACF decays
essentially monotonically and converges to the VACF of a (Markovian) system with delta-distributed friction kernel.
In between these cases, one may observe VACFs with a ``plateau" at intermediate times. Similar behavior occurs in the presence of colored noise (see next paragraph).

We finally consider the MSDs in the underdamped (white-noise-)systems with center topology, which are indicated in Fig.~\ref{fig:VACFmap} by dash-dotted green lines. At very short times $\Delta t$, we observe ballistic behavior ($\propto (\Delta t)^2$)
in all cases considered, consistent with the appearance of the inertial term in Eq.~(\ref{eq:markovall_v}). Further, the long-time behavior of the MSD is always diffusive, (i.e., $\propto\Delta t$). In contrast,
at intermediate times, the MSD may display changes in the curvature (white region) and even non-monotonicity, as the plot at largest $k_2b_2$ (see top left) reveals. 
Inspecting Fig.~\ref{fig:VACFmap} we find that such complex behavior of the MSD occurs together with oscillations in the VACF. 
\subsubsection{Systems with colored noise}
\begin{figure}
\begin{center}
	\includegraphics[scale=.9]{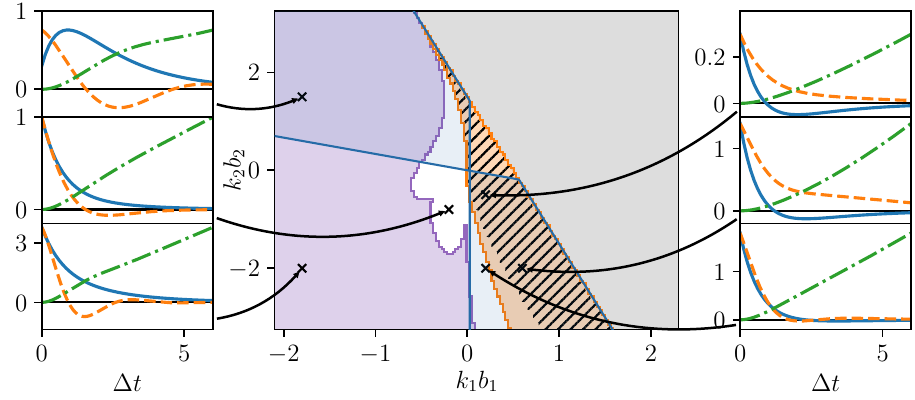}
    \caption{Same as Fig.~\ref{fig:VACFmap}, but with colored noise ($T_1=T_2=T_0$). To enhance visibility, the VACFs and MSDs are scaled
    (with factors given in Table~\ref{tab:mapScales} in the Appendix)}.
	\label{fig:VACFmap_colored}
	\end{center}
 \end{figure}

So far we have focused on the white-noise case, $T_1=T_2=0$. To illustrate the impact of colored noise, $\xi_c$ (which represents an additional type of memory),
we show in Fig.~\ref{fig:VACFmap_colored} the analog of the map presented in Fig.~\ref{fig:VACFmap}, yet with $T_1=T_2=T_0$.
We recall that the VACF now involves both $c_V^\text{I}(\Delta t)$ and $c_V^\text{II}(\Delta t)$ stemming from the correlations of the colored noise
[see Eq.~(\ref{eq:vacfInt})], while the friction kernels remain unchanged.
Interestingly, the differences regarding the VACFs are rather subtle (apart from different magnitudes), as a comparison
with Fig.~\ref{fig:VACFmap} reveals.
Closer inspection shows that there are slight shifts of the boundaries separating different regions of the maps.
More prominently, the MSDs in the case $\xi_c\neq 0$ exhibit a smoother behavior 
at the coupling parameters, where changes of the curvature occur at $\xi_c=0$ (compare, e.g., left bottom panels of Fig.~\ref{fig:VACFmap} and Fig.~\ref{fig:VACFmap_colored}).
We may conclude, however, that the main trends observed already in the white-noise system, where memory arises
solely through the friction kernel, remain unchanged.

\begin{figure}
\begin{center}
		\includegraphics[scale=.85]{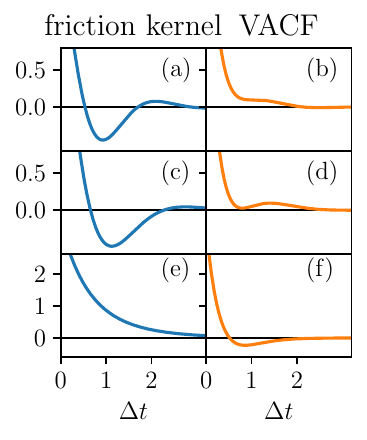}
    \caption{Examplary friction kernels [(a), (c), (e)] and corresponding VACFs [((b), (d), (f)] for underdamped systems with colored noise that fulfill the FDR.
	The parameters of the underlying (fully coupled) Markovian system are given in Table~\ref{tab:underdampedFDR}.}
	\label{fig:corr_underdampedFDR}
	\end{center}
 \end{figure}

To close this section, we consider in Fig.~\ref{fig:corr_underdampedFDR} examples of correlation functions in systems with all-to-all coupling of the variables
and colored noise.
The results do not indicate any profound changes in the VACF relative to the VACFs already discussed for simpler topologies or in the absence of colored noise.
A special feature of the cases considered in Fig.~\ref{fig:corr_underdampedFDR}, however, is that the friction kernel and the colored
noise are related. This will be discussed in more detail in the next section.
\section{\label{sec:fdr}Fluctuation-Dissipation relation}
The non-Markovian equations derived in Sec.~\ref{sec:projection} formally resemble the so-called generalized Langevin equations (GLE) resulting from Mori-Zwanzig projection \cite{zwanzig1961memory,mori1965transport}
of, e.g., the Hamilton equations
of a many-particle system onto a slow variable (such as the velocity of a tagged particle). For systems in thermal equilibrium, the frictional and stochastic forces that result from integrating out
irrelevant variables balance each other, as expressed via the fluctuation-dissipation relation (FDR) \cite{zwanzig2001nonequilibrium} (sometimes referred to as fluctuation-dissipation theorem of second kind)
between the friction kernel and the (colored) noise correlations. Here we explore the presence of a FDR for both, the under- and overdamped version of our model.
However, the validity of FDR alone does not guarantee
thermal equilibrium in a strict sense, different from the detailed balance (DB) condition \cite{RevModPhys.48.571,2312} which acts on the level of the underlying Markovian equations. For the (simpler) case of overdamped systems,
we therefore study both measures (for a recent discussion of the FDR and DB in underdamped models, see \cite{PhysRevE.101.022120,loos2020thermodynamic}).

\subsection{\label{sec:fdr_under}Underdamped case}
In the underdamped case, the relation of interest is given by
\begin{align}
		\langle \xi(t)\xi(t+\Delta t)\rangle \stackrel{?}{=}
		\Gamma(\Delta t)\langle v^2\rangle,
		\label{eq:FDR}
\end{align}
where the left hand side represents the correlations of the total noise, $\xi=\xi_0+\xi_\text{c}$, and $\Gamma$ is the full friction kernel including the term involving $\gamma_0$ (compare Eq.~(\ref{eq:solutionxAllCoupled})).
Using (\ref{eq:gammaplusminus}), (\ref{eq:frictionKernelUnder}), (\ref{eq:noisecorrplusminus}) and (\ref{eq:noiseCorrelation}), 
and noting the de-coupling between white and colored noise, i.e., $\langle\xi_0\xi_c\rangle=0$ (which follows from the
statistical independency of $\xi_0$ and $\xi_i$), Eq.~(\ref{eq:FDR}) can be written as
\begin{align}
2\gamma_0k_BT_0\delta(\Delta t)
	+\langle \xi_\text{c}(t)\xi_\text{c}(t+\Delta t)\rangle
~	\overset{\text{?}}{=}~\langle v^2\rangle \big[ 2 \gamma_0 \delta(\Delta t) 
+ \gamma^+(\Delta t)+\gamma^-(\Delta t)\big]~.
	\label{eq:FDR2}
\end{align}
A valid FDR implies that Eq.~(\ref{eq:FDR2}) is fulfilled at {\em all} times $\Delta t$.
As the structure of Eq.~(\ref{eq:FDR2}) reveals, this hangs essentially on two aspects. 

First, the prefactors of the delta distributions appearing on both sides of Eq.~(\ref{eq:FDR2}) 
must be identical (a difference can not be compensated by the remaining functions, which are finite sums of exponentials). In an {\em equilibrium} underdamped system,
the equipartition theorem (with mass $m=1$) dictates $\langle v^2\rangle =c_V(t=0)=k_BT_0$, such that the equality is trivially fulfilled. 
However, equipartition is {\em not} automatically fulfilled in the underdamped systems at hand. We note in passing that we only consider the equipartition theorem for the variable $v$ here, and neglected the variables $y_1$, $y_2$
involved in the underlying Markovian model Eq.~(\ref{eq:markovallNoMatrix}). The argument is that $y_1$, $y_2$ are considered as auxiliary variables which do not necessarily have a physical meaning [see the discussion
below Eqs.~(\ref{eq:markovallNoMatrix})]. 

Using Eq.~(\ref{eq:vacfInt}) for the VACF, 
the condition that the equipartition theorem for the velocity $v$ holds can be expressed as
\begin{align}
	\langle v^2\rangle &= c_V(0)=\tilde c_V^\text{I}k_B T_0\gamma_0+\tilde c_V^\text{II.1}k_BT_1+\tilde c_V^\text{II.2}k_BT_2
	~\stackrel{!}{=} ~k_BT_0~.
	\label{eq:eqpUnderdamped}
\end{align}
where $\tilde c_V^\text{I}\equiv c_V^\text{I}(0)/\gamma_0k_BT_0$ and $\tilde c_V^\text{II,i}\equiv c_V^\text{II}(0)/k_BT_i$ ($i=1,2$). The definition of $c_V^\text{II}(0)$ follows directly from Eq.~(\ref{eq:vacfInt}) 
when one notes that the noise correlations entering the integral separate into contributions proportional to $T_1$ and $T_2$, respectively.

The second condition for the FDR (\ref{eq:FDR2}) to hold is the equality of the contribution from the remaining colored noise and friction kernel in Eq.~(\ref{eq:FDR2}) {\em at any time}.
Clearly, for (time-)non-local friction kernels $\gamma^\pm$, the FDR at finite $\Delta t$ can only hold if the correlations 
of the colored noise are non-zero. Therefore, systems with $T_i=0$ (such as considered in Fig.~\ref{fig:VACFmap}) automatically violate the FDR.
There is a further, rather general observation which can be made: the colored-noise correlations have their maximum at $\Delta=0$ [as can be seen explicitly from Eq.~(\ref{eq:noisecorrplusminus})]. Therefore, the FDR can only be fulfilled if the non-trivial part of the friction kernel has its maximum at $\Delta t=0$ as well. This immediately implies that the systems with ring topology break the FDR.

To check the FDR at finite times, we recall from Eqs.~(\ref{eq:noiseCorrelation}) and (\ref{eq:noisecorrplusminus}) and (\ref{eq:frictionKernelUnder}) that both functions depend on time through
the exponentials $\exp(-(a_1+a_2\pm\omega)t/2)$. The FDR can thus only be fulfilled if
the coefficients of these two exponentials match. This leads to a two-dimensional matrix equation for the vector $(T_1,T_2)$ 
\begin{align}
	\begin{pmatrix}
		\eta_1^+-\tilde c_V^\text{II.1}g^+ &\eta_2^+ -\tilde c_V^\text{II.2} g^+\\
		\eta_2^--\tilde c_V^\text{II.1} g^-&\eta_2^- -\tilde c_V^\text{II.2} g^-
	\end{pmatrix}
	\begin{pmatrix}
	T_1\\ T_2
	\end{pmatrix}=
	\tilde c_V^\text{I}\gamma_0 T_0
	\begin{pmatrix}
		g^+\\ g^-
	\end{pmatrix}~. 
	\label{eq:FDRtogether}
\end{align}
where [from Eq.~(\ref{eq:noiseCorrelation})], $\eta_i^{\pm}=c_i^+ c_i^+/(a_1+a_2\pm\omega)+c_i^+c_i^-/(a_1+a_2)$, $i=1,2$, $g^{\pm}$ is defined according to Eq.~(\ref{eq:gplusminus}), 
and we have replaced $\langle v^2\rangle$ by Eq.~(\ref{eq:eqpUnderdamped}).

From Eq.~(\ref{eq:FDRtogether}) it is seen that the validity of the FDR depends on both
the coupling parameters (determining the entries of the matrices) and on the temperatures of the auxiliary
variables $T_1$ and $T_2$. To test the FDR for fixed coupling parameters and fixed $T_0$, we solve
Eq.~(\ref{eq:FDRtogether}) with respect to the vector $(T_1,T_2)$ (thereby inverting the matrix on the left hand side).
If the resulting products $T_1$ and $T_2$ are real and positive, then the FDR can be satisfied. 
These values automatically also satisfy the equipartition condition, Eq.~(\ref{eq:eqpUnderdamped}).

\begin{figure}
\begin{center}
	\includegraphics[scale=0.8]{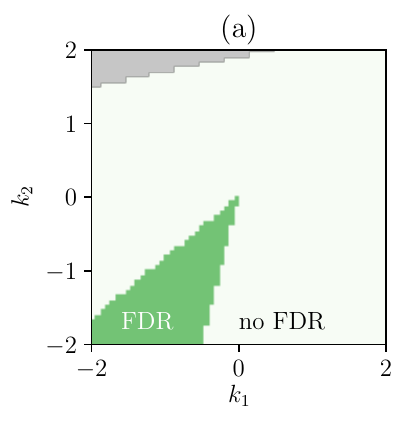}
	\includegraphics[scale=0.8]{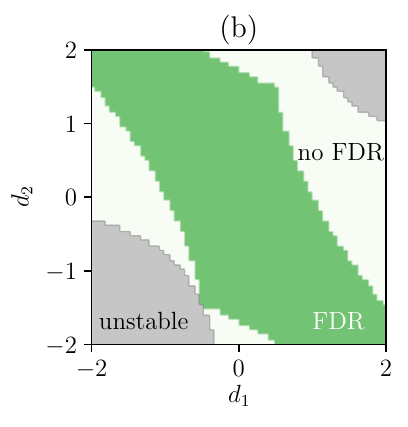}   
	\caption{Illustration of coupling parameter regions where the FDR and the equipartition theorem for the colloid's velocity in the underdamped system can be fulfilled (green areas). The corresponding (positive) temperatures $T_1$, $T_2$ depend on the parameter combination considered. In the grey areas, the system is unstable. (a) parameter plane spanned by $k_1$ and $b_1$ (at fixed $d_1=-2=-d_2$, $b_1=b_2=1$), 
	(b) parameter plane spanned by $d_1$ and $d_2$ (at fixed $k_1=k_2=-0.5$, $b_1=b_2=0.5$). The remaining parameters in both parts are $T_0=1$, $\gamma_0=1$, $a_1=0.5$, $a_2=1.5$. }
	\label{fig:regions_underdampedFDR}
	\end{center}
 \end{figure}

Due to the large parameter space, it is essentially impossible to test the validity of the FDR for every combination
of coupling constant and friction parameters in the underlying Markovian model. However, some observations can be made.
For $d_1=d_2=0$ (center topology), closer analysis of Eq.~(\ref{eq:FDRtogether}) reveals that positive temperatures
$T_1$, $T_2$ can only be found if
$k_ib_i<0$. For all-to-all coupled systems, we moreover find (from a numerical analysis) as a major trend that the ``back-and-forth" coupling constants
must have opposite signs and a similar magnitude, i.e., $k_i\approx -b_i$, $d_1\approx - d_2$ (for an illustration, see Fig.~\ref{fig:regions_underdampedFDR}).
This signals that non-reciprocal coupling is a necessary ingredient for the FDR to hold (along with different bath temperatures). At first sight, this finding may seem somewhat surprising, since one may suspect
a violation of Newton's third (action-reaction) law. However, one has to keep in mind that
the auxiliary variables do not necessarily have a physical interpretation; moreover, if one assigns them a physical character, the latter may be velocity-like rather than position-like.
Therefore, ``non-reciprocal" here does not necessarily imply unbalanced mechanical forces. We also note that a similar observation, i.e., non-reciprocal coupling as a condition for a valid FDR, has been made in 
a (micro-)rheological model with one auxiliary variable ($n=1$) proposed in \cite{raikher2010theory}, as well as in the $n=1$-case of the linear underdamped model studied in \cite{PhysRevE.101.022120} (see Appendix~F of that paper).

Finally, we note that the question
of whether the FDR is fulfilled is {\em not} easily inferred from inspection of the VACF and the MSD. 
To give some examples, we show in the right part of Fig.~\ref{fig:corr_underdampedFDR} three friction kernels with
corresponding VACFs, where the FDR is satisfied. Further, in parts (a) and (b), the underlying Markovian network is all-to-all coupled, yielding
the possibility of oscillatory friction kernels. It is seen that the corresponding VACF displays a plateau or oscillations.
Part (c) is an example from a network with center topology, where the friction kernel decays  monotonically and the VACF 
displays a minimum. This case is similar to the behavior in the white region in Fig.~\ref{fig:VACFmap_colored}.
 \subsection{\label{sec:fdr_over}Overdamped case}
In an overdamped system, it is natural to assume that at the times considered, the velocities have relaxed to their equilibrium values. Thus, the equipartition theorem 
for the velocity $\langle v^2\rangle=k_BT_0$ (with $m=1$)
is fulfilled automatically, 
and the FDR related to the non-Markovian equation~(\ref{eq:GLEoverdamped}) is given by 
\begin{align}
		\langle \xi(t)\xi(t+\Delta t)\rangle \stackrel{?}{=}
		\tilde\Gamma(\Delta t)k_BT_0
		\label{eq:FDR_over}
\end{align}
or (using the cancellation of delta-like terms)
\begin{align}
\langle \xi_\text{c}(t)\xi_\text{c}(t+\Delta t)\rangle
	\overset{\text{?}}{=} k_BT_0\big[ 
 \gamma_\text{o}^+(\Delta t)+\gamma_\text{o}^-(\Delta t)\big]
	\label{eq:FDR2_over}
\end{align}
where the functions on the right side are defined in Eq.~(\ref{eq:frictionOverdamped}). Equation~(\ref{eq:FDR2_over}) can again be written as a matrix equation for the temperatures $T_1$, $T_2$, which can be solved numerically.

\begin{figure}
	\centering
		\includegraphics[width=0.7\textwidth]{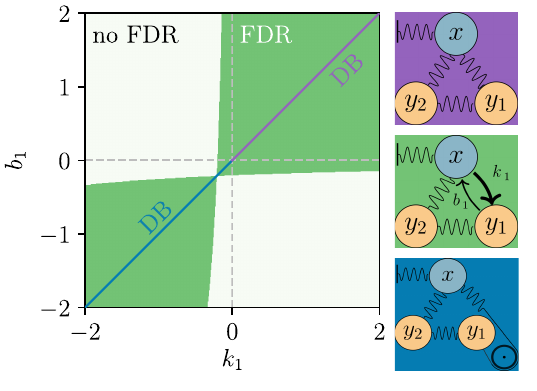}
	\caption{Illustration of the ranges of the coupling constants $k_1$, $b_1$ in which
	the FDR can be fulfilled in the overdamped case (green areas). The corresponding (positive) temperatures $T_1$, $T_2$ depend on the combination ($k_1,b_1$) considered, while the remaining parameters are fixed ($T_0=1$, $k_2=b_2=1$, $d_1=d_2=1$, and $a_0 = 3.6$, $a_1=3$, $a_2=2$). Along the
	diagonal line (reciprocal coupling), detailed balance (DB) holds. For $k_1=b_1>0$ (violet part of the diagonal), the parameters correspond to
	a Brownian particle coupled to other Brownian particles and a fixed point via springs, see top panel on the right. For $k_1=b_1<0$ (blue part of the diagonal), the system can still fulfill DB (mechanically,
	$y_1$ may be viewed as a deflection pulley), see bottom panel on the right. The middle panel on the right illustrates the case of non-reciprocal coupling between $x$ and $y_1$, where however, the FDR is still fulfilled.}
	\label{fig:overdampedFDR}
\end{figure}

A result of such a calculation is presented in Fig.~\ref{fig:overdampedFDR}, where we have fixed all coupling constants except $k_1$, $b_1$ which relate the main variable (here $x$) and the first auxiliary variable, $y_1$. 
In the green, shaded areas of the $k_1$-$b_1$-plane, the calculated temperatures $T_1$ and $T_2$ are real and positive and thus, the FDR holds. 
Interestingly, these areas
include large regions characterized by non-reciprocal coupling ($k_1\neq b_1$) and even
(small) regions where $k_1$ and $b_1$ have opposite signs.
Moreover, we find (numerically) that in the areas where the FDR holds, the potential energy in the harmonic trap fulfills equipartition, i.e., $1/2\langle x^2\rangle a_0' = 1/2 \, k_BT_0$.

A special situation occurs on the diagonal, i.e., for the reciprocal choice $k_1$=$b_1$ (note that the other coupling constants are reciprocally chosen as well). Here, the resulting temperatures are all equal, i.e., $T_1=T_2=T_0$. Moreover, this fully reciprocal situation satisfies not only the FDR, but also the more severe detailed balance (DB) condition \cite{RevModPhys.48.571}. The latter can be checked by considering the probability currents (related to the three variables) in the corresponding 
Fokker-Planck equation, see~\cite{loos2020thermodynamic,RevModPhys.48.571,PhysRevE.101.022120}. We here interpret the auxiliary variables $y_i$ as position-like variables, i.e., they are even under time-reversal.
It is important to note that, under certain conditions, the DB condition for the linear system at hand can also be fulfilled with non-reciprocal coupling or different temperatures \cite{loos2020thermodynamic}. In the case considered in Fig.~\ref{fig:overdampedFDR}, however, DB only holds on the diagonal.
Combining this with our findings regarding the FDR, we see that there are indeed large parameter regions (with non-reciprocal coupling, see green areas),
where DB is violated, but the FDR is satisfied. 
This signals that (non-)equilibrium measures based on the non-Markovian model alone must be treated with care, consistent with Ref.~\cite{Crisanti2012}. In particular,~\cite{Crisanti2012} explicitly shows that even when the entropy production is considered, expressions that are based on the non-Markovian equation alone may lead to vanishing entropy production in non-equilibrium states, contray to one's expectation.

The regions in Fig.~\ref{fig:overdampedFDR} can be analysed on the Markovian level of our model, as well. Along the diagonal $k_i=b_i$, the Markovian 
model corresponds to overdamped Brownian particles coupled to each other with springs (see top panel on the right). For negative $k_i=b_i$, the force resulting from the spring has the opposite 
direction, as for positive $k_i=b_i$, which corresponds to the use of a deflection pulley (see bottom panel). For $k_i\neq b_i$, the coupling is non-reciprocal and does not correspond to any type of mechanical coupling (see middle panel).

\section{\label{sec:appl}Application}
In this final section, we discuss an example illustrating that the rather generic models and corresponding correlation functions
introduced before can be used to provide a simple description (fitting) of correlation functions in a real system.

\begin{figure}
	\centering
	\includegraphics[width=.80\textwidth]{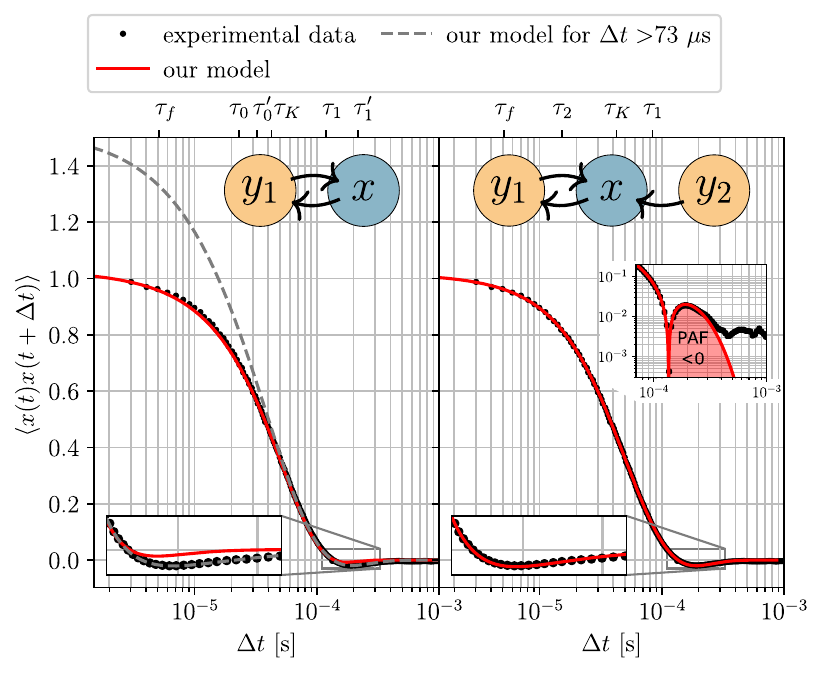}
	\includegraphics[width=.1755\textwidth]{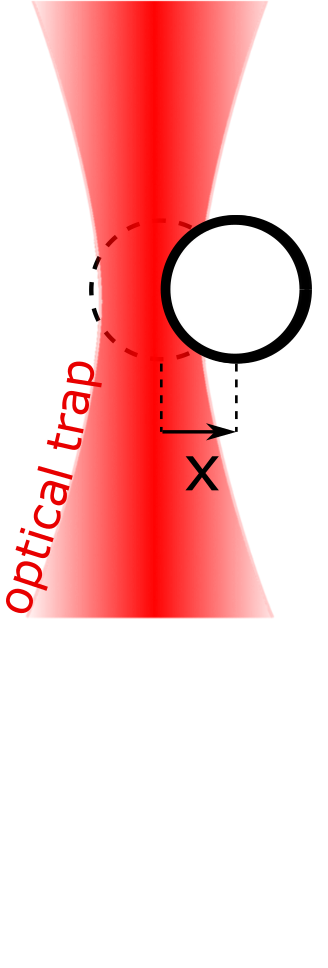}
	\caption{Positional autocorrelation function (PACF) of a colloidal bead in a static optical trap.
		Black dots denote normalized experimental data \cite{franosch2011resonances}.
		The negative overshoot at intermediate times arises due to hydrodynamic backflow at the
		timescale $\tau_f$ of vortices
		generated at the surface of the bead. Left panel: Two fits based on the overdamped model with a single auxiliary variable ($y_1$) bidirectionally coupled to $x$ as indicated by the inserted schematic. The fits either approximate the 
		initial decay (solid red line fitted to all data) or 
		anticorrelations 
		(dashed grey line, only data from $\Delta t>73.4\,\mu s$). Right panel: Fit based on a overdamped model with two auxiliary variables $y_1$, $y_2$, where the coupling between $x$ and $y_2$ is unidirectional. In both panels, we have included insets
		with an enlargement of the overshoot region. The right panel includes additionally a double-logarithmic representation of that region. Right to the main panel we show a schematic of the colloid (white disc) 
confined to an optical trap. 
The variable $x$ corresponds to the distance to the center of the trap.}
	\label{fig:fit}
\end{figure}

Specifically, inspired by the combined experimental-theoretical study presented in Ref.~\cite{franosch2011resonances}, 
we consider correlation functions of a micron-sized colloidal particle suspended in a viscous solvent.
The colloidal particle is optically ``trapped", i.e., exposed to a confining potential. The focus of Ref.~\cite{franosch2011resonances}
was the measurement and theoretical description of colored noise related to ``anticorrelations" in the position autocorrelation function (PACF), that is, regions
of negative values of the PACF at intermediate times, as visible from the dotted line in Fig.~\ref{fig:fit}. We note that the experimental data of the log-scaled PACF have been 
extracted from \cite{franosch2011resonances} (PDF-Version) manually in the time regimes of interest, using the tool {\em WebPlotDigitizer} \cite{rohatgi2011webplotdigitizer}.

Physically, the anticorrelations are induced
by hydrodynamic backflow: the particle transfers momentum to the surrounding 
medium \cite{franosch2011resonances}, which generates vortices at the sphere's surface. These vortices then diffuse over one particle radius on a timescale $\tau_f$,
which eventually leads to a backward force acting on the particle at later times \cite{kheifets2014observation}.

Clearly, the simple models introduced in the present paper cannot provide a full hydrodynamic description of this effect (as it was done, e.g.,
in \cite{PhysRevA.2.2005,fodor2015generalized}), in particular, our models can not describe the detailed time-dependency of 
the PACF ($\propto t^{-3/2}$) in the long-time limit.
Still, by a systematic fit of the experimental data for the PACF in \cite{franosch2011resonances} we found that the {\em general shape} of the PACF, particularly the emergence of anticorrelations,
can be mimicked by our simple linear model with two auxiliary variables. Since the effects we are mainly
interested in take place well-above the
ballistic timescale, we employ, for the sake
of simplicity, the overdamped model~(\ref{eq:markovover}). Note that the parameter $a_0$ (coupling linearly to the position)
then corresponds to the harmonic trap.

Our fitting strategy was guided by the following considerations. Despite the simple structure of the full system of overdamped equations~(\ref{eq:markovover}), it comes with
a rather large set of parameters. Clearly, one is interested in the model with the least complexity, i.e., the minimum number of parameters. 
In the first place, we have thus attempted to fit the data with a reduced version of Eq.~(\ref{eq:markovover}) involving only one auxiliary variable (i.e., $n=1$), $y_1$, which is subject to a heat bath. This set-up, which involves five parameters
($a_0$, $a_1$, $T_0$, $T_1$ and the product $k_1b_1$) yields single-exponential memory and colored noise.
%
The corresponding PACF is calculated by repeating the steps outlined in Sec.~\ref{sec:correlations} and \ref{sec:green} for the case of $n=1$.
To obtain the optimal parameters we have employed a least square fitting scheme based on the Nelder Mead algorithm 
\cite{nelder1965simplex}.
One issue of this scheme is that it finds a local minimum of the residuals 
without anticorrelations in the PACF. 
We overcome this problem by starting the fit with a truncated data set ($\Delta t>73\mu s$),
where mostly the anticorrelations are included and values for small $\Delta t$ 
are left out.  Then, a new fit with more data is started (using the result of the previous fit) until all data are included.

The left panel of Fig.~\ref{fig:fit} shows our best attempts to resemble the experimental data
with this $n=1$ fitting approach.
Specifically, the dashed line in Fig.~\ref{fig:fit} (left) 
has been obtained with a truncated data set, using the parameters 
$a_0=$\num{3.09e4}s$^{-1}$, $a_1=$\num{4.62e3}s$^{-1}$, $k_1b_1=$\num{-1.4e+08}s$^{-1}$,
$\langle \xi_0^2(t)\rangle=$\num{8.09e+04}s$^{-1}$ and 
$\langle \xi_1^2(t)\rangle=$\num{3.78e+04}s$^{-1}$.
The solid line has been obtained 
by continuing the scheme above and eventually using all data points, yielding 
$a_0=$\num{4.34e4}s$^{-1}$, $a_1=$\num{8.36e3}s$^{-1}$, $k_1b_1=$\num{-5.66e+08}s$^{-1}$,
$\langle \xi_0^2(t)\rangle=$\num{2.70e+04}s$^{-1}$ and 
$\langle \xi_1^2(t)\rangle=$\num{4.21e+04}s$^{-1}$.
As Fig.~\ref{fig:fit} (left) reveals, one can either mimic the initial decay
(solid line) 
{\em or} the decay of the anticorrelations
(dashed line), but not the full PACF. The problem persists if one uses
different initial parameters and truncated data sets.

The  failure of the fitting procedure with $n=1$ indicates that (at least) two auxiliary variables
are required. In this context we note that the definition of a Markovian ``model" corresponding to given data is not
unique, and we here aim for a most simple one.
 To keep the number of fitting parameters as small as possible, we fixed $a_0$ to
$a_0^{-1}=\tau_K=42.7$ $\mu$s, which is the timescale corresponding to the harmonic force given in the experiment
(i.e., $\tau_K^{-1}=K/\gamma_S$, where $K$ and $\gamma_S$ are the trap stiffness and (Stokes) friction coefficient taken from \cite{franosch2011resonances}).
The auxiliary variables were incorporated 
using a special, non-reciprocal, version of the center topology (where $y_2$ is coupled only unidirectionally to $x$, i.e., $b_2=0$) 
with positive temperatures $T_i$. With $a_2>0$ and $T_2>0$, the variable $y_2$ is then governed by an Ornstein-Uhlenbeck process.
The resulting non-Markovian model involves a (single-) exponentially decaying kernel [see Eq.~(\ref{eq:gammastar})], with decay time $\tau_1=1/a_1$; $y_2$ does not contribute to 
the friction kernel but to the colored noise. 

The resulting PACF is shown as a red line in the right panel of
Fig.~\ref{fig:fit}. The parameters (beyond $a_0$ given above) read 
	$a_1=$\num{1.19e4}s$^{-1}$,
	$a_2=$\num{6.55e4}s$^{-1}$,
	$k_1b_1=$\num{-4.96e+07}s$^{-1}$,
	$k_2=$\num{-3.62e3}s$^{-1}$,
	$\langle \xi_0^2(t)\rangle=$\num{1.86e+04}s$^{-1}$,
	$\langle \xi_1^2(t)\rangle=$\num{1.65e+04}s$^{-1}$, and
$\langle \xi_2^2(t)\rangle=$\num{1.43e+07}s$^{-1}$. One can see that it describes the measured data quite well.
It is interesting to look in more detail at the timescales (marked at the top of Fig.~\ref{fig:fit}). The smallest timescale is $\tau_f$, which was determined in \cite{franosch2011resonances}  and measures the vortex diffusion.
%
%
%
The other three timescales correspond to the three variables of our model. We recall that $\tau_K$ was used to fix $a_0$.
Further, $\tau_2=1/a_2$, which is much smaller than $\tau_K$, is located within the region of initial decay.
This conforms with the idea
that the Ornstein-Uhlenbeck process $y_2$ provides a ``kick" into the system affecting the first decay. 
Indeed, due to the unidirectional coupling between $x$ and $y_2$ our model has purely non-equilibrium character in the sense that it does not satisfy the FDR, nor DB \cite{loos2020thermodynamic}.
This is in contrast to the (underdamped) Langevin equation used in Ref.~\cite{franosch2011resonances} where the FDR is assumed to hold from the outset.
The largest timescale $\tau_1=1/a_1$, which is related to $y_1$, is larger by several orders of magnitude than $\tau_K$. It falls in the region where anti-correlations occur. Thus, in this sense, the coupling with the auxiliary variable $y_1$ mimics the effect of the hydrodynamic backflow.
 
\begin{figure}
	\centering
	\includegraphics{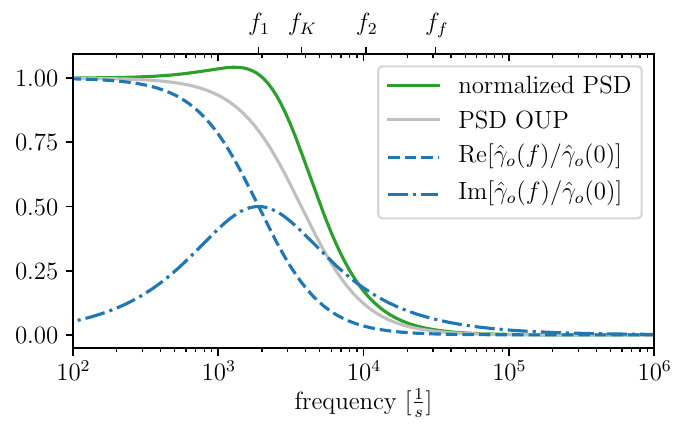}
	\caption{
	Power spectral density (PSD) of the PACF and the Fourier-transformed memory kernel (real and imaginary part) corresponding to model with $n=2$. The normalized
PSD (green line) shows a peak around $f_1 = a_1/2\pi$.
The real part of the friction kernel decays 
monotonically (dashed line), while the imaginary part shows a peak at $f_1$ (dash-dotted
line). Also shown is the PSD resulting from a simple Ornstein-Uhlenbeck process (silver line),
to which our model reduces upon decoupling $x$ from the auxiliary variables.}
	\label{fig:PSD}
\end{figure}

Finally, we consider in Fig \ref{fig:PSD} the correlation and memory function 
in frequency space (obtained by Fourier-transformation of the curves shown in 
Fig (\ref{fig:fit}). 
Considering the spectrum enables us to see oscillatory modes which,
in turn, provides another interesting perspective on the dynamics of the 
colloidal particle in the harmonic trap. 
To get a better impression of the effect of the memory,
we have supplemented the plot by the spectrum of the corresponding Markovian 
process without memory, that is, the relaxation
	of an overdamped particle in 
the harmonic trap which is a usual Ornstein-Uhlenbeck process (grey line).
Recall that our model includes this as a special case when we decouple the 
auxiliary variables $y_{1,2}$ from $x$. 
In comparison, the spectrum of the particle subject to memory forces 
(i.e., with coupling to $y_{1,2}$) resembles the behavior of the 
Ornstein-Uhlenbeck process in the limits of high and low frequencies.
However, it has an additional pronounced peak close to $f_1=a_1/2\pi$,
indicating oscillatory modes with and around this specific frequency.
As it was shown in \cite{franosch2011resonances}, this peak is the result of 
resonance between the hydrodynamic memory and the overdamped relaxation in 
the trap, which induces anticorrelations in the PACF and positional oscillations.
As we see here, the non-trivial memory induced by the coupling to two auxiliary 
variables likewise creates resonant 
oscillations, similar to hydrodynamic memory.
Interestingly, the normalized imaginary part of the memory kernel is nonzero in this 
very frequency range and likewise exhibits a peak at the frequency $f_1$,
which is associated with the variable $y_1$. 
This can be seen from the Fourier transform of Eqs.~(\ref{eq:gammastar}) and (\ref{eq:frictionOverdamped}), namely $\hat \gamma_o(f)=k_1b_1/[f_1(f_1-if)4\pi^2]$.
Analogously to our observations in time domain, this again suits to the idea 
that the variable $y_1$ imitates the effect of the hydrodynamic memory in this frequency range.
\section{\label{sec:outlook}Conclusions and outlook}
In this paper we have provided a comprehensive study of autocorrelation functions of non-Markovian systems which can be mapped to linear Markovian systems with two embedded (auxiliary) variables. 
The two levels of description can be connected
by a projection procedure, very similar to Mori-Zwanzig-like coarse-graining approaches. In the resulting equation for the variable
of interest, the non-Markovianity then appears through (time-nonlocal) friction kernels and colored noise. Contrary to the standard GLE, however,
here we do not assume that the coupling terms involved in the Markovian equations are reciprocal, i.e., derivable from an underlying Hamiltonian.  In other words,
the resulting ``forces" (related to position-dependent terms) can violate Newton's third law.
Our analysis pertains to linear systems with few variables, which enables us to perform all calculations analytically. This includes, in particular, the calculation of autocorrelation
functions of the velocity (underdamped systems) or position (overdamped systems), which are experimentally accessible quantities in a large variety of soft matter systems.

As a first step, we have explored in detail the interplay between the coupling topology of the underlying Markovian equations and the resulting friction kernel. In particular, already with two auxiliary variables
one can describe highly non-trivial kernels including oscillations, finite values at zero time, and non-monotonicity with maxima or minima at finite times. For the linear systems at hand,
these kernels are independent of the heat baths to which the auxiliary variables are coupled.
Second, already the relatively simple center topology with (white) noise only in the main variable
allows to describe a range of behaviors of time-dependent autocorrelation functions of the velocity (underdamped case) or position (overdamped) case,
including, in particular, oscillatory behavior (see Fig.~\ref{fig:VACFmap}). The first minimum of the autocorrelation function can be related to the instantaneous value of the friction kernel.

In calculating the correlation functions, the case of non-zero auxiliary temperatures ($T_1\neq 0$, $T_2\neq 0$) and thus, colored noise, turned out to be doable, yet quite cumbersome.
Therefore it was interesting to see that including this colored noise did not introduce truly new features in the correlation functions. Thus, when modelling friction kernels
by simple ansatzes, starting from deterministic auxiliary equations seems to be the most plausible choice. However, colored noise becomes crucial when one requires that the non-Markovian model satisfies the FDR (or even DB). Clearly, this is well established
for standard GLEs, but not in the present systems, where the existence of equilibrium is not guaranteed. We have separately investigated the FDR in the underdamped and overdamped systems, thereby deriving explicit conditions which were then checked
numerically.
In the actual underdamped systems considered,
the FDR can only be satisfied for {\em non-reciprocal} coupling. This is different in the overdamped case where, for the topology considered, the coupling (corresponding to mechanical forces) had to be reciprocal for the FDR to be fulfilled.
In the overdamped case, we have additionally checked the DB condition. Interestingly, we found parameter regions (characterized by non-reciprocal coupling) where FDR is fulfilled but not DB. 
We note that for the class of systems studied here (no additional external forces), situations where FDR is fulfilled but DB is violated do not occur if only one auxiliary variable is involved \cite{loos2020thermodynamic}. This confirms (along with our results for the correlation functions) that the choice $n=2$ introduces indeed novel aspects of the system's properties.

The results of the present paper may be of interest in several contexts. This concerns, first, the modelling of actual experimental data. A classical example is passive or active microrheology, where one measures the diffusive (or active) motion
of a micro-sized object through a viscous fluid. An example of such a modelling was discussed in Sec.~\ref{sec:appl} where we demonstrated that our model with two auxiliary variables
provides indeed a good description of the experimentally measured PACF data. Furthermore, the simple ansatzes suggested here may also be used to fit numerical data (e.g., from Molecular dynamics or Brownian Dynamics
simulations) for autocorrelation functions, where the corresponding kernels then follow from the fitted coupling constants. For example, a two-exponential kernel (with a maximum) has recently found to be sufficient to explain
oscillatory PACF in a colloidal system \cite{berner2018oscillating}. Our results for the appearance of oscillations are consistent with \cite{berner2018oscillating}, and by exploring a broader parameter space (i.e., topologies of the underlying Markovian network), we provide a complete picture of the achievable behaviors. Moreover, the simple models may
also be used as an inspiration to introduce, at a relatively low computational cost, memory effects in numerical simulations of complex many-particle systems by adding few degrees of freedom to each true particle (for a recent discussion of the use of embedding techniques in numerical simulations, see \cite{jung2018generalized}).

From a more fundamental point of view, the present study contributes to the understanding of non-Markovian systems regarding their non-equilibrium character and, connected with that, the impact on measurable quantities. Here we have focused on the FDR and DB as measures of equilibrium
(for a discussion of further thermodynamic notions in related linear systems, see, e.g., \cite{PhysRevE.101.022120,loos2020thermodynamic}). Interestingly, we have found that the autocorrelation functions do not really reflect whether or not these equilibrium conditions are fulfilled. Finally, while the present work was mostly devoted to the non-Markovian dynamics of a single particle, it seems interesting to explore an extension towards coupled systems. Work in this direction is on the way.

\section*{Acknowledgements}
This work was funded by the Deutsche Forschungsgemeinschaft (DFG, German Research Foundation) - Projektnummer 163436311 - SFB 910. 
We thank Felix H{\"o}fling and Arthur Straube for inspiring discussions regarding correlation functions in Fourier space. 
 \appendix
 \section{\label{sec:derivationProjection}Projection}
 In this Appendix we provide the details of the derivation of the non-Markovian equations in Sec.~\ref{sec:projection}.
 \subsection{\label{sec:derivationKernel}Derivation of kernel and colored noise}
 To project the Markovian Langevin equations (\ref{eq:markovallNoMatrix}) and (\ref{eq:markovover})
 onto equations for the main dynamical variables,
 we utilize the Laplace transform \cite{kuhfittig1978laplace} defined by
 $\mathcal{L}\{g(t)\}(s)=\int_0^\infty g(t)e^{-st}dt\equiv \hat g(s)$. In the underdamped case, we obtain
 from Eq.~(\ref{eq:markovallNoMatrix}) 
 \begin{subequations}
	\label{eq:allLaplace}
\begin{align}
  s \hat v-v_0 &= -\gamma_0 \hat v + k_1\hat v+k_2\hat y_2+\hat\xi_0
  \label{eq:shatx}\\
  s \hat y_1 &= -a_1 \hat y_1 + b_1\hat v+d_1\hat y_2+\hat\xi_1
  \label{eq:shaty1}\\
  s \hat y_2 &= -a_2 \hat y_2 + b_2\hat v+d_2\hat y_1+\hat\xi_2~.
  \label{eq:shaty2}
\end{align}
\end{subequations}
Here, $v_0$ is the initial velocity, and we have set the initial values of
$y_1$ and $y_2$ to zero. This is since we are mainly interested in the long-time behavior, where the system has achieved a steady state, where the initial conditions become irrelevant.
Rearranging and inserting the equations into one another we find
%
\begin{align}
    s\hat v-v_0 =& -\gamma_0 \hat v
    +\left[\frac{b_1(s+a_2)}{(s+a_1)(s+a_2)-d_1d_2}+\frac{d_1b_2}{(s+a_1)(s+a_2)-d_1d_2}\right]k_1\hat v\nonumber\\
    &+\left[\frac{b_2(s+a_1)}{(s+a_2)(s+a_1)-d_1d_2}+\frac{d_2b_1}{(s+a_1)(s+a_2)-d_1d_2}\right]k_2\hat v\nonumber\\
    &+\left[\frac{k_1(s+a_2)}{(s+a_1)(s+a_2)-d_1d_2}\hat\xi_1+\frac{k_1d_1}{(s+a_1)(s+a_2)-d_1d_2}\hat\xi_2\right]\nonumber\\
    &+\left[\frac{k_2(s+a_1)}{(s+a_2)(s+a_1)-d_1d_2}\hat\xi_2+\frac{k_2d_2}{(s+a_1)(s+a_2)-d_1d_2}\hat\xi_1\right]~.
    \label{eq:Laplacestep}
\end{align}
To transform Eq.~(\ref{eq:Laplacestep})  back into the time domain we use the
convolution theorem $\mathcal{L}^{-1}\{\hat f(s)\cdot \hat g(s)\}=\int_0^tf(t-u)g(u)du$ \cite{kuhfittig1978laplace}
and the inverse Laplace transforms
	\begin{align}
		\mathcal{L}^{-1}\left[\frac{1}{(s+a_1)(s+a_2)-d_1d_2}\right]
		&=\frac{2}{\omega} \e^{-(a_1+a_2)t/2}\text{Sinh}\frac{\omega t}{2}
		\nonumber\\
	\mathcal{L}^{-1}\left[\frac{s+a_2}{(s+a_1)(s+a_2)-d_1d_2}\right]
	&=\e^{-(a_1+a_2)t/2}\left[\text{Cosh}\frac{\omega t}{2}
	+\frac{a_2-a_1}{\omega}\text{Sinh}\frac{\omega t}{2}\right],
	\label{eq:Laplacespecial}
	\end{align}
where $\omega$ is given in Eq.~(\ref{eq:omega}). Applying the relations~(\ref{eq:Laplacespecial}) to the terms
on the right side of Eq.~(\ref{eq:Laplacestep}) and sorting with respect to $v$ finally
brings us to Eq.~(\ref{eq:solutionxAllCoupled}), which involves the colored noise
\begin{align}
		\xi_c(t)=&\int_0^t dt' \exp(-\omega^+t') \left[c_1^+
		\xi_1(t')+c_2^{+}\xi_2(t')\right]
		\nonumber\\
&+\int_0^t dt' \exp (-\omega^{-}t')  \left[c_1^-
		\xi_1(t')+c_2^{-}\xi_2(t')\right]
\label{eq:coloredNoise}
\end{align}
with the constants
\begin{align}
c_1^{\pm} =\frac{k_1}{2}\mp\frac{a_2k_1-a_1k_1+2d_2k_2}{2\omega};~~~
c_2^{\pm} =\frac{k_2}{2}\mp\frac{a_1k_2-a_2k_2+2d_1k_1}{2\omega},
\label{eq:c12pm}
\end{align}
and
\begin{align}
	\omega_{\pm}=\frac{a_1+a_2\pm \omega}{2}~.
\label{eq:omegapm}
\end{align}
%
%

 \subsection{\label{sec:derivationNoisecorr}Correlations of the colored noise}
To derive Eq.~(\ref{eq:noiseCorrelation}) in the main text, we multiply the instantaneous values of the colored noise,
$\xi_c$ [see Eq.~(\ref{eq:coloredNoise})] at two times $t$, $t+\Delta t$ and then perform the noise average.
Taking into account that the noise terms related to {\em different} auxiliary variables are uncorrelated,
i.e., $\langle \xi_1(t)\xi_2(t')\rangle=0$ $\forall t,t'$, it is clear that the final correlation function is a sum of contributions
involving $\xi_1$ and $\xi_2$, respectively.
Here we focus on the terms related to  $\xi_1$. Assuming $\Delta t>0$, one finds
	\begin{align}
		\langle \xi_c(t+\Delta t)\xi_c(t)\rangle_1
		=&\int_0^t dt' \int_0^{t+\Delta t}du~\Big(
		~c_1^+c_1^+e^{-\omega_+t'-\omega_+u}
		+c_1^+c_1^-e^{-\omega_+t'-\omega_-u}\nonumber\\
		&+c_1^-c_1^+e^{-\omega_-t'-\omega_+u}
		+c_1^-c_1^-e^{-\omega_-t'-\omega_-u}\Big)
		\langle \xi_1(t-t')\xi_1(t+\Delta t-u)\rangle\\
		=&2k_BT_1\int_0^tdt' c_1^+c_1^+e^{-\omega_+(2t'+\Delta t)}
		+c_1^-c_1^+ e^{-(\omega_+ +\omega_-)t'-\omega_-\Delta t}\nonumber \\
		&+c_1^+c_1^- e^{-(\omega_+ +\omega_-)t'-\omega_+\Delta t}
		+c_1^-c_1^-e^{-\omega_-(2t'+\Delta t)}\\
		=&-2k_BT_1\Big[
		\frac{c_1^+c_1^+}{2\omega_+}e^{-\omega_+(2t'+\Delta t)}
		+\frac{c_1^-c_1^+}{\omega_++\omega_-} e^{-(\omega_+ +\omega_-)t'-\omega_-\Delta t}
		\nonumber \\
		&+\frac{c_1^+c_1^-}{\omega_++\omega_-}e^{-(\omega_+ +\omega_-)t'-\omega_+\Delta t}
	+\frac{c_1^-c_1^-}{2\omega_-}e^{-\omega_-(2t'+\Delta t)}\Big]_{t'=0}^t\\
		=&2k_BT_1\Big(-
		\frac{c_1^+c_1^+}{2\omega_+}e^{-\omega_+(2t+\Delta t)}
		-\frac{c_1^-c_1^+}{\omega_++\omega_-} e^{-(\omega_+ +\omega_-)t-\omega_-\Delta t}
		\nonumber \\
		&\hspace{1cm}-\frac{c_1^+c_1^-}{\omega_++\omega_-}e^{-(\omega_+ +\omega_-)t-\omega_+\Delta t}-\frac{c_1^-c_1^-}{2\omega_-}e^{-\omega_-(2t+\Delta t)}
		\nonumber \\
	&+
		\frac{c_1^+c_1^+}{2\omega_+}e^{-\omega_+\Delta t}
		+\frac{c_1^-c_1^+}{\omega_++\omega_-} e^{-\omega_-\Delta t}
		+\frac{c_1^+c_1^-}{\omega_++\omega_-}e^{-\omega_+\Delta t}
	+\frac{c_1^-c_1^-}{2\omega_-}e^{-\omega_-\Delta t}\Big)\\
		=&2k_BT_1
		\Big[
\frac{c_1^+c_1^+}{2\omega_+}e^{-\omega_+\Delta t}\left(1-e^{-w_+ 2t}\right)
+\frac{c_1^-c_1^+}{\omega_++\omega_-} e^{-\omega_-\Delta t}\left(1-e^{-(w_++\omega_-) t}\right)\nonumber \\
&+\frac{c_1^+c_1^-}{\omega_++\omega_-}e^{-\omega_+\Delta t}\left(1-e^{-(w_++\omega_-) t}\right)
+\frac{c_1^-c_1^-}{2\omega_-}e^{-\omega_-\Delta t}\left(1-e^{-w_- 2t}\right)\Big]~.
		\label{<+label+>}
	\end{align}
By the same steps, one obtains the contribution involving $\xi_2$, yielding the
same result but with changed indices $1\leftrightarrow 2$. The sum of the two contributions
then yields the full correlation function of the colored noise, Eq.~(\ref{eq:noiseCorrelation}).
\subsection{\label{sec:derivationKernelover}Friction kernel in the overdamped case}
For the overdamped case, the derivation of the projected equation~(\ref{eq:xProjectedG}) is identical to that of 
Eq.~(\ref{eq:solutionxAllCoupled}). In order to find the friction kernel for the velocity $\dot x$, we use integration by parts 
\begin{align}
	\int_0^tG(t-t')x(t')dt
	=&-2g^{+}\,\frac{e^{-(a_1+a_2+\omega)t/2}x(0)-x(t)}{a_1+a_2+\omega} \nonumber\\
	&
  -2g^{-}\,\frac{ e^{-(a_1+a_2-\omega)t/2}x(0)-x(t)}{a_1+a_2-\omega}
+\int_0^t\tilde{\Gamma}(t-t')v(t')dt'~,
	\label{eq:integrationByParts}
\end{align}
where $\tilde{\Gamma}(t)$ is the desired friction kernel defined below Eq.~(\ref{eq:GLEoverdamped}) in the main text,
and $g^{\pm}$ has been defined in Eq.~(\ref{eq:gplusminus}).
In the long time limit, the exponential functions in the first lines of (\ref{eq:integrationByParts})
vanish. What remains from these terms is a term proportional to the position, which we
name $-(a_0'-a_0)x(t)$. This finally leads to Eq.~(\ref{eq:GLEoverdamped}).

\section{\label{sec:kernelParameters}Parameters}
In this Appendix we provide numerical values characterizing various parameter sets used in this study.
Table~\ref{tab:kernel} contains parameters for the non-trivial parts of the friction kernel, $\gamma(t)$, plotted in Fig.~\ref{fig:kernel}.
\begin{table}
\caption{\label{tab:kernel}Parameters for the functions $\gamma(t)$ plotted in Fig.~\ref{fig:kernel}.
The abbreviations are ``dec" for monotonically decaying (grey lines), ``max" (``min") for non-monotonic with a maximum (minimum)
at a finite time (blue solid (dashed) lines), and ``osc" for oscillating (red lines).}
	\begin{tabular}{|l|c|c|c|c|c|c|c|c|}
\hline
	  & $a_1$ & $a_2$ & $k_1$ & $k_2$ & $b_1$ & $b_2$ & $d_1$ & $d_2$\\
\hline
Fig. 2(a) & & & & & & & &\\
dec & 2& 2& -1 & -1 & 1 & 1 & 1 & 1 \\
max          & 5.37& 0.79& -2 & -3.35 & -1.01 & 0.29 & 0.04 & 0.61 \\
osc          & 0.85& 0.29& -0.069 & 0.58 & -2.03 & 0.84 & -1.54 & 2.88 \\
\hline
Fig. 2(b) & & & & & & & &\\
min & 2.18& 2.13& 0 & -1.34 & -0.82 & 0 & 0 & 2.05 \\
 max         & 3.64& 0.79& 0 & 0.05 & 32.17 & 0 & 0 & -3.25 \\
\hline
Fig. 2(c)  & & & & & & & &\\
dec & 1& 1& -1 & 0 & 1 & 0 & 1 & 0.10 \\
 osc         & 0.10& 0.20& -0.67 & 0 & 1 & 0 & -1 & 1 \\
min          & 2& 0.20& -1.0 & 0 & 1 & 0 & -0.68 & 1 \\
\hline
Fig. 2(d)   & & & & & & & &\\
dec & 1& 2& -0.5 & -0.50 & 1 & 1 & 0 & 0 \\
min          & 0.50& 1.50& 0.95 & -1.90 & 1 & 1 & 0 & 0 \\
 max         & 0.50& 1.50& -1.44 & 1.44 & 1 & 1 & 0 & 0 \\
\hline
	\end{tabular}
\end{table}
Table~\ref{tab:mapScales} provides the scaling factors of the VACFs and MSDs in Fig.~\ref{fig:VACFmap} and Fig.~\ref{fig:VACFmap_colored}, which have been introduced to enhance visibility.
\begin{table}
	\caption{Factors by which the VACFs and MSDs plotted in Fig.~\ref{fig:VACFmap}  
		and Fig.~\ref{fig:VACFmap_colored} are scaled:}
	\label{tab:mapScales}
	\centering
	\begin{tabular}{|c|c|c|}
		\hline
		Figure \ref{fig:VACFmap} & & \\
		panel & VACF & MSD \\
		\hline
		upper left & 1.58 & 0.74 \\middle left & 2.64 & 0.56 \\lower left & 12.25 & 11.80 \\upper right & 0.64 & 0.06 \\middle right & 3.76 & 0.49 \\lower right & 5.35 & 1.21 \\
\hline
	\end{tabular}
	\begin{tabular}{|c|c|c|}
		\hline
		Figure \ref{fig:VACFmap_colored} & & \\
panel & VACF & MSD \\
\hline upper left & 0.48 & 0.09 \\middle left & 2.28 & 0.41 \\lower left & 5.19 & 1.36 \\upper right & 0.57 & 0.05 \\middle right & 1.68 & 0.13 \\lower right & 3.03 & 0.44 \\
\hline
	\end{tabular}
\end{table}
Table~\ref{tab:underdampedFDR} contains the parameters for the systems considered in Fig.~\ref{fig:corr_underdampedFDR}.
\begin{table}
	\caption{Parameters for the functions $\gamma(t)$ and corresponding VACFs plotted in 
	Fig.~\ref{fig:corr_underdampedFDR}.}
	\label{tab:underdampedFDR}
	\centering
	\begin{tabular}{|c|cccccccccccc|}
		\hline
		& $T_0$ & $T_1$ & $T_2$ & $\gamma_0$ & $a_1$ & $a_2$ & $k_1$ & $k_2$ & $b_1$ & $b_2$ & $d_1$ & $d_2$ \\
		\hline
	a) & 3.72 & 2.80 & 1.10 & 4 & 1 & 2 & -0.83 & -1.22 & 1 & 1  & 3 & -2.50\\
	b) & 3.72 & 3.06 & 0.44 & 4 & 1 & 2 & -1.22 & -2.00 & 1 & 1  & 2 & -2.00 \\
	c) & 3.72 & 0.58 & 0.93 & 4 & 1 & 2 & -1.60 & -2.00 & 1 & 1  & 0 & 0.00 \\
\hline
	\end{tabular}
\end{table}
\section{Calculation of the velocity autocorrelation function based on the Green's function}
\label{sec:green}
Our analytical calculation of the VACF, $c_V$, defined in Eq.~(\ref{eq:vacf})
relies on a Fourier transform of Eq.~(\ref{eq:solutionxAllCoupled}) (thus taking advantage of the convolution theorem) 
and subsequent back transformation, yielding
\begin{align}
  v(t)= \int_{-\infty}^\infty ds \lambda(s)\left(\xi_0(t-s)+\xi_\text{c}(t-s)\right)~,
\label{eq:xtimeconv}
\end{align}
where $\lambda(s)$ is the Green's Function in the time domain. Its Fourier transform, $\hat\lambda(f)$ (with $f$ being the frequency), follows from the transform of Eq.~(\ref{eq:solutionxAllCoupled}) as
\begin{align}
\hat\lambda(f)=-\frac{1}{if-\hat \Gamma(f)+i\left(\kappa/f\right)}.
\label{eq:lambdaFourier}
\end{align}
The explicit calculation of $\lambda(s)$ thus involves the calculation of the Fourier transform of the full friction kernel, $\hat\Gamma(f)$, as well as the inverse transformation
of $\hat\lambda(f)$. In the following we show how this can be done using a partial fraction decomposition (for an alternative approach, see \cite{PhysRevE.72.061107}). 
We focus on the case $\kappa=0$ (i.e., no confining potential), but briefly comment on the case $\kappa>0$ at the end of this Appendix.

As a first step, we calculate the Fourier transform of the full friction kernel, $\Gamma(t)$. Using 
$\hat h=\mathcal{F}[h(t)](f)=\int_{-\infty}^\infty dt h(t)\exp[ift]$ and inserting Eq.~(\ref{eq:gammaplusminus})
we find
\begin{align}
  \hat \Gamma (f)=
  \int_{0}^\infty & dt ~ \left[2\gamma_0 \delta(t)
  +g^+\exp\left(-(a_1+a_2+\omega)\frac{t}{2}\right)\right. \nonumber \\
  &\left.+g^-\exp\left(-(a_1+a_2-\omega)\frac{t}{2}\right)
  \right] \times\exp\left[ift\right]
  \end{align}
where we have omitted the negative part of the integral because $\Gamma(t)=0$, $t<0$ (due to causality), and
$g^\pm$ and $\omega$ are defined in Eqs.~(\ref{eq:gplusminus}) and (\ref{eq:omega}), respectively.
Carrying out the integrals yields
  \begin{align}
  \hat\Gamma (f)&= \gamma_0 + \frac{g^+}{\frac{1}{2}(a_1+a_2+\omega)-if}
+ \frac{g^-}{\frac{1}{2}(a_1+a_2-\omega)-if}~.
\label{eq:khatintegral} 
\end{align}

Inserting Eq.~(\ref{eq:khatintegral}) into Eq.~(\ref{eq:lambdaFourier})
and using Eq.~(\ref{eq:omegapm}), we obtain 
\begin{align}
\hat\lambda(f)&=-\left[if-\gamma_0 - \frac{g^+}{\omega^+-if}\right.
 \left.-\frac{g^-}{\omega^--if}\right]^{-1}
 \label{eq:hatLambda}
\end{align}
To perform the back transformation we perform a partial fraction decomposition. To this end we find the common denominator of the fraction appearing
in the brackets $[\ldots ]$ in Eq.~(\ref{eq:hatLambda}) and substitute $\tilde  f = i f$. This yields
\begin{align}
	\hat \lambda (-i \tilde  f) =&\frac{ -(\omega^+-\tilde  f)(\omega^--\tilde  f)
	}{(a_0+\tilde f)(\omega^+-\tilde f)(\omega^--\tilde f)
		+g^+(\omega^--\tilde f)+g^-(\omega^+-\tilde f)}~.
    \label{eq:lhatexpand}
  \end{align}
 
 The denominator is a polynomial of third order, which therefore has three complex roots $ F_1, F_2$ and $ F_3$.
 These can be expressed explicitly in terms of the parameters $a_i, k_i$ and $b_i$. The most convenient way to find these
 (quite cumbersome) expressions is to use computer algebra programs, for example Mathematica.

Having obtained the roots, Eq.~(\ref{eq:lhatexpand}) can be rewritten as 
\begin{align}
\hat\lambda( f)&=-\frac{(\omega^+-i f)(\omega^--i f)}{(i f- F_1)(i f- F_2)(i f- F_3)},
\label{eq:partial1}
\end{align}
which we now aim to express as
\begin{align}
\hat \lambda ( f) = \frac{\epsilon_1}{i f- F_1}+\frac{\epsilon_2}{i f- F_2}+\frac{\epsilon_3}{i f- F_3}~.
\label{eq:lhatPartial}
\end{align}
The remaining task is to determine the constants $\epsilon_1,\epsilon_2$ and $\epsilon_3$. 
To this end we obtain the common demoninator in Eq.~(\ref{eq:lhatPartial}) and compare the resulting numerator
with the one in Eq.~(\ref{eq:partial1}), yielding
\begin{align}
-(\omega^+-i f)(\omega^--i f) \overset{!}{=}&~ ~\epsilon_1 (i f- F_2)(i f- F_3)
+\epsilon_2(i f- F_1)(i f- F_3)
\nonumber \\&
+\epsilon_3(i f- F_1)(i f- F_2)~.
\label{eq:findctilde}
\end{align}
The left and right hand sides of Eq.~(\ref{eq:findctilde}) are polynomials of degree two in $f$.
In order to fulfil (\ref{eq:findctilde}) for all values of $ f$, the coefficients of the polynomials on both sides have to be identical.
This yields the following explicit expressions for $\epsilon_1$, $\epsilon_2$, and $\epsilon_3$ in terms of the roots,
\begin{align}
\epsilon_1 &= \frac{(\omega^+ -  F_1) ( F_1-\omega^- )}{( F_1 -  F_2) ( F_1 -  F_3)};~
\epsilon_2 = \frac{(\omega^+ -  F_2) (\omega^- -   F_2)}{( F_1 -  F_2) ( F_2 -  F_3)};~
\epsilon_3 = \frac{(\omega^+ -  F_3) (\omega^- -   F_3)}{( F_1 -  F_3) (- F_2 + F_3)}~.
\label{eq:calcd}
\end{align}

The last step of the calculation is the back transform of Eq.~(\ref{eq:lhatPartial}) into the time domain,
\begin{align}
\lambda(t)&=\sum_{k=1}^3\mathcal{F}^{-1}\left[\frac{\epsilon_k}{if-F_k}\right]
=-\sum_{k=1}^3 \epsilon_k \exp[-F_kt]\theta(t)
\label{eq:backtrafo}
\end{align}
In Eq.~(\ref{eq:backtrafo}), the Heaviside step function ensures causality. This function arises if we assume that 
the real parts of $F_k$ are positive, which we (numerically) find to be satisfied for all parameters where the Markovian system is stable
(see discussion of Fig.~\ref{fig:VACFmap}). This finally leads to 
\begin{align}
  \lambda(t) &=- 
  \sum_{k=1}^3 \epsilon_k e^{-F_kt},
	\label{eq:lambdareal}
\end{align}
where the coefficients $\epsilon_k$ are given in Eq.~(\ref{eq:calcd}) and $F_k$ denote the poles of $\hat\lambda$. In Eq.~(\ref{eq:lambdareal}) we have assumed
that all real parts are positive 
to ensure causality (i.e., $\lambda(t)=0$ for $t<0$) \cite{sekimoto2010stochastic}.

We are now in the position to calculate the VACF.
 Combining Eqs.~(\ref{eq:vacf}) and
 (\ref{eq:xtimeconv}) we obtain 
\begin{align}
  c_V(\Delta t)&=\int_{-\infty}^t ds\int_{-\infty}^{t+\Delta t} ds'~ \lambda(s)\lambda(s')
  \Big\langle \xi(t-s)\xi(t+\Delta t-s')\Big\rangle,
\end{align}  
where $\lambda(s)$ is given explicitly in Eq.~\ref{eq:lambdareal}) and $\xi(t)=\xi_0(t)+\xi_\text{c}(t)$. Multiplying out the noise correlations we arrive at Eq.~(\ref{eq:vacfInt}) in the main text.

Finally, in the case $\kappa>0$ [corresponding to a finite confining potential in Eq.~(\ref{eq:markovall_v})],  the denominator's degree of Eq.~(\ref{eq:partial1}) is 
increased by one (yielding four roots), and we end up with an expression for $\lambda(t)$ identical to Eq.~(\ref{eq:backtrafo}),
except that the sum contains four (instead of three) terms. We have used this approach to calculate
the VACF for $\kappa>0$ for systems with center topology.

\section*{References}
\bibliography{main.bib}

\begin{thebibliography}{10}

\bibitem{raikher2010theory}
Yu~L Raikher and VV~Rusakov.
\newblock Theory of {B}rownian motion in a {J}effreys fluid.
\newblock {\em Journal of Experimental and Theoretical Physics},
  111(5):883--889, 2010.

\bibitem{berner2018oscillating}
Johannes Berner, Boris M{\"u}ller, Juan~Ruben Gomez-Solano, Matthias
  Kr{\"u}ger, and Clemens Bechinger.
\newblock Oscillating modes of driven colloids in overdamped systems.
\newblock {\em Nat. Commun.}, 9(1):1--8, 2018.

\bibitem{gotze2008complex}
Wolfgang G{\"o}tze.
\newblock {\em Complex dynamics of glass-forming liquids: A mode-coupling
  theory}, volume 143.
\newblock OUP Oxford, 2008.

\bibitem{nagai2015collective}
Ken~H Nagai, Yutaka Sumino, Raul Montagne, Igor~S Aranson, and Hugues
  Chat{\'e}.
\newblock Collective motion of self-propelled particles with memory.
\newblock {\em Phys. Rev. Lett.}, 114(16):168001, 2015.

\bibitem{PhysRevLett.121.078003}
N~Narinder, Clemens Bechinger, and Juan~Ruben Gomez-Solano.
\newblock Memory-induced transition from a persistent random walk to circular
  motion for achiral microswimmers.
\newblock {\em Phys. Rev. Lett.}, 121:078003, Aug 2018.

\bibitem{kursten2017giant}
R{\"u}diger K{\"u}rsten, Vladimir Sushkov, and Thomas Ihle.
\newblock Giant kovacs-like memory effect for active particles.
\newblock {\em Phys. Rev. Lett.}, 119(18):188001, 2017.

\bibitem{marchetti2013hydrodynamics}
M~Cristina Marchetti, Jean-Fran{\c{c}}ois Joanny, Sriram Ramaswamy,
  Tanniemola~B Liverpool, Jacques Prost, Madan Rao, and R~Aditi Simha.
\newblock Hydrodynamics of soft active matter.
\newblock {\em Rev. {M}od. {P}hys.}, 85(3):1143, 2013.

\bibitem{narinder2018memory}
N~Narinder, Clemens Bechinger, and Juan~Ruben Gomez-Solano.
\newblock Memory-induced transition from a persistent random walk to circular
  motion for achiral microswimmers.
\newblock {\em Phys. {R}ev. {L}ett.}, 121(7):078003, 2018.

\bibitem{aguilar2018critical}
Daniel Aguilar-Hidalgo, Steffen Werner, Ortrud Wartlick, Marcos
  Gonz{\'a}lez-Gait{\'a}n, Benjamin~M Friedrich, and Frank J{\"u}licher.
\newblock Critical point in self-organized tissue growth.
\newblock {\em Phys. Rev. Lett.}, 120(19):198102, 2018.

\bibitem{mitterwallner2020non}
Bernhard~G Mitterwallner, Christoph Schreiber, Jan~O Daldrop, Joachim~O
  R{\"a}dler, and Roland~R Netz.
\newblock Non-{M}arkovian data-driven modeling of single-cell motility.
\newblock {\em Phys. Rev. E}, 101(3):032408, 2020.

\bibitem{lange2006generalized}
Oliver~F Lange and Helmut Grubm{\"u}ller.
\newblock Generalized correlation for biomolecular dynamics.
\newblock {\em Proteins: Structure, Function, and Bioinformatics},
  62(4):1053--1061, 2006.

\bibitem{carmele2019non}
Alexander Carmele and Stephan Reitzenstein.
\newblock Non-{M}arkovian features in semiconductor quantum optics: quantifying
  the role of phonons in experiment and theory.
\newblock {\em Nanophotonics}, 8(5):655--683, 2019.

\bibitem{loos2019heat}
Sarah~AM Loos and Sabine~HL Klapp.
\newblock Heat flow due to time-delayed feedback.
\newblock {\em Sci. Rep.}, 9(1):1--11, 2019.

\bibitem{khadka2018active}
Utsab Khadka, Viktor Holubec, Haw Yang, and Frank Cichos.
\newblock Active particles bound by information flows.
\newblock {\em Nat. Commun.}, 9(1):1--9, 2018.

\bibitem{khadem2019delayed}
Seyed Mohsen~Jebreiil Khadem and Sabine~HL Klapp.
\newblock Delayed feedback control of active particles: a controlled journey
  towards the destination.
\newblock {\em Phys. Chem. Chem. Phys.}, 21(25):13776--13787, 2019.

\bibitem{scholl2016control}
Eckehard Sch{\"o}ll, Sabine~HL Klapp, and Philipp H{\"o}vel.
\newblock {\em Control of self-organizing nonlinear systems}.
\newblock Springer, 2016.

\bibitem{straube2020rapid}
Arthur~V Straube, Bartosz~G Kowalik, Roland~R Netz, and Felix H{\"o}fling.
\newblock Rapid onset of molecular friction in liquids bridging between the
  atomistic and hydrodynamic pictures.
\newblock {\em Communications Physics}, 3(1):1--11, 2020.

\bibitem{klages2008anomalous}
Rainer Klages, G{\"u}nter Radons, and Igor~M Sokolov.
\newblock {\em Anomalous transport: foundations and applications}.
\newblock John Wiley \& Sons, 2008.

\bibitem{hofling2013anomalous}
Felix H{\"o}fling and Thomas Franosch.
\newblock Anomalous transport in the crowded world of biological cells.
\newblock {\em Reports on Progress in Physics}, 76(4):046602, 2013.

\bibitem{gernert2016feedback}
Robert Gernert, Sarah~AM Loos, Ken Lichtner, and Sabine~HL Klapp.
\newblock Feedback control of colloidal transport.
\newblock In {\em Control of Self-Organizing Nonlinear Systems}, pages
  375--392. Springer, 2016.

\bibitem{lichtner2010feedback}
Ken Lichtner and Sabine~HL Klapp.
\newblock Feedback-controlled transport in an interacting colloidal system.
\newblock {\em EPL (Europhysics Letters)}, 92(4):40007, 2010.

\bibitem{Kyaw_2020}
Thi~Ha Kyaw, Victor~M. Bastidas, Jirawat Tangpanitanon, Guillermo Romero, and
  Leong-Chuan Kwek.
\newblock Dynamical quantum phase transitions and non-{M}arkovian dynamics.
\newblock {\em Phys. Rev. A}, 101(1), Jan 2020.

\bibitem{munakata2014entropy}
T~Munakata and ML~Rosinberg.
\newblock Entropy production and fluctuation theorems for {L}angevin processes
  under continuous non-{M}arkovian feedback control.
\newblock {\em Phys. Rev. Lett.}, 112(18):180601, 2014.

\bibitem{loos2020thermodynamic}
Sarah A.~M. Loos and Sabine H.~L. Klapp.
\newblock Thermodynamic implications of non-reciprocity.
\newblock {\em arXiv preprint arXiv:2008.00894}, 2020.

\bibitem{debiossac2020thermodynamics}
Maxime Debiossac, David Grass, Jose~Joaquin Alonso, Eric Lutz, and Nikolai
  Kiesel.
\newblock Thermodynamics of continuous non-{M}arkovian feedback control.
\newblock {\em Nat. Commun.}, 11(1):1--6, 2020.

\bibitem{di2020thermodynamic}
Ivan Di~Terlizzi and Marco Baiesi.
\newblock A thermodynamic uncertainty relation for a system with memory.
\newblock {\em arXiv preprint arXiv:2005.05226}, 2020.

\bibitem{zwanzig1961memory}
Robert Zwanzig.
\newblock Memory effects in irreversible thermodynamics.
\newblock {\em Phys. Rev.}, 124(4):983, 1961.

\bibitem{mori1965transport}
Hazime Mori.
\newblock Transport, collective motion, and {B}rownian motion.
\newblock {\em Progress of theoretical physics}, 33(3):423--455, 1965.

\bibitem{daldrop2017external}
Jan~O Daldrop, Bartosz~G Kowalik, and Roland~R Netz.
\newblock External potential modifies friction of molecular solutes in water.
\newblock {\em Phys. Rev. X}, 7(4):041065, 2017.

\bibitem{metzler2017gaussianity}
Ralf Metzler.
\newblock Gaussianity fair: the riddle of anomalous yet non-gaussian diffusion.
\newblock {\em Biophysical journal}, 112(3):413, 2017.

\bibitem{wang2012brownian}
Bo~Wang, James Kuo, Sung~Chul Bae, and Steve Granick.
\newblock When {B}rownian diffusion is not gaussian.
\newblock {\em Nature materials}, 11(6):481--485, 2012.

\bibitem{chechkin2017brownian}
Aleksei~V Chechkin, Flavio Seno, Ralf Metzler, and Igor~M Sokolov.
\newblock {B}rownian yet non-gaussian diffusion: from superstatistics to
  subordination of diffusing diffusivities.
\newblock {\em Phys. Rev. X}, 7(2):021002, 2017.

\bibitem{Yag58}
A.M. Yaglom.
\newblock Correlation theory of processes with random stationary nth
  increments.
\newblock {\em Am. Math. Soc. Transl.}, 8:87--141, 1958.

\bibitem{jeonvivi}
Jae-Hyung Jeon, Vincent Tejedor, Stas Burov, Eli Barkai, Christine
  Selhuber-Unkel, Kirstine Berg-S{\o}rensen, Lene Oddershede, and Ralf Metzler.
\newblock In vivo anomalous diffusion and weak ergodicity breaking of lipid
  granules.
\newblock {\em Phys. Rev. Lett.}, 106(4):048103, 2011.

\bibitem{szymanski2009elucidating}
Jedrzej Szymanski and Matthias Weiss.
\newblock Elucidating the origin of anomalous diffusion in crowded fluids.
\newblock {\em Phys. Rev. Lett.}, 103(3):038102, 2009.

\bibitem{jung2018generalized}
Gerhard Jung, Martin Hanke, and Friederike Schmid.
\newblock Generalized {L}angevin dynamics: construction and numerical
  integration of non-{M}arkovian particle-based models.
\newblock {\em Soft matter}, 14(46):9368--9382, 2018.

\bibitem{siegle2010markovian}
Peter Siegle, Igor Goychuk, Peter Talkner, and Peter H{\"a}nggi.
\newblock {M}arkovian embedding of non-{M}arkovian superdiffusion.
\newblock {\em Phys. Rev. E}, 81(1):011136, 2010.

\bibitem{kappler2019non}
Julian Kappler, Victor~B Hinrichsen, and Roland~R Netz.
\newblock Non-{M}arkovian barrier crossing with two-time-scale memory is
  dominated by the faster memory component.
\newblock {\em The European Physical Journal E}, 42(9):119, 2019.

\bibitem{zwanzig2001nonequilibrium}
Robert Zwanzig.
\newblock {\em Nonequilibrium statistical mechanics}.
\newblock Oxford University Press, 2001.

\bibitem{Shankar2018}
Suraj Shankar and M~Cristina Marchetti.
\newblock Hidden entropy production and work fluctuations in an ideal active
  gas.
\newblock {\em Phys. Rev. E}, 98(2):020604(R), 2018.

\bibitem{Caprini2019}
Lorenzo Caprini, Umberto Marini~Bettolo Marconi, Andrea Puglisi, and Angelo
  Vulpiani.
\newblock The entropy production of {O}rnstein--{U}hlenbeck active particles: a
  path integral method for correlations.
\newblock {\em J. Stat. Mech. Theor. Exp.}, 2019(5):053203, 2019.

\bibitem{Dabelow2019}
Lennart Dabelow, Stefano Bo, and Ralf Eichhorn.
\newblock Irreversibility in active matter systems: Fluctuation theorem and
  mutual information.
\newblock {\em Phys. Rev. X}, 9(2):021009, 2019.

\bibitem{Martin2020}
David Martin, J{\'e}r{\'e}my O'Byrne, Michael~E Cates, {\'E}tienne Fodor,
  Cesare Nardini, Julien Tailleur, and Fr{\'e}d{\'e}ric van Wijland.
\newblock Statistical {M}echanics of {A}ctive {O}rnstein {U}hlenbeck
  {P}articles.
\newblock {\em arXiv preprint arXiv:2008.12972}, 2020.

\bibitem{PhysRevE.101.022120}
Roland~R. Netz.
\newblock Approach to equilibrium and nonequilibrium stationary distributions
  of interacting many-particle systems that are coupled to different heat
  baths.
\newblock {\em Phys. Rev. E}, 101:022120, Feb 2020.

\bibitem{loos2019fokker}
Sarah A.~M. Loos and Sabine H.~L. Klapp.
\newblock {F}okker-{P}lanck equations for time-delayed systems via {M}arkovian
  embedding.
\newblock {\em J. Stat. Phys.}, 177:95--118, 2019.

\bibitem{franosch2011resonances}
Thomas Franosch, Matthias Grimm, Maxim Belushkin, Flavio~M Mor, Giuseppe Foffi,
  L{\'a}szl{\'o} Forr{\'o}, and Sylvia Jeney.
\newblock Resonances arising from hydrodynamic memory in {B}rownian motion.
\newblock {\em Nature}, 478(7367):85, 2011.

\bibitem{kowalik2019memory}
Bartosz Kowalik, Jan~O Daldrop, Julian Kappler, Julius~CF Schulz, Alexander
  Schlaich, and Roland~R Netz.
\newblock Memory-kernel extraction for different molecular solutes in solvents
  of varying viscosity in confinement.
\newblock {\em Phys. Rev. E}, 100(1):012126, 2019.

\bibitem{doi:10.1063/1.4928456}
Antonio Lasanta and Andrea Puglisi.
\newblock An itinerant oscillator model with cage inertia for mesorheological
  granular experiments.
\newblock {\em The Journal of Chemical Physics}, 143(6):064511, 2015.

\bibitem{PhysRevLett.114.198001}
Camille Scalliet, Andrea Gnoli, Andrea Puglisi, and Angelo Vulpiani.
\newblock Cages and anomalous diffusion in vibrated dense granular media.
\newblock {\em Phys. Rev. Lett.}, 114:198001, May 2015.

\bibitem{baldovin2019langevin}
Marco Baldovin, Andrea Puglisi, and Angelo Vulpiani.
\newblock Langevin equations from experimental data: The case of rotational
  diffusion in granular media.
\newblock {\em PloS one}, 14(2):e0212135, 2019.

\bibitem{PhysRevE.102.012908}
Andrea Plati and Andrea Puglisi.
\newblock Slow time scales in a dense vibrofluidized granular material.
\newblock {\em Phys. Rev. E}, 102:012908, Jul 2020.

\bibitem{hansen1990theory}
Jean-Pierre Hansen and Ian~R McDonald.
\newblock {\em Theory of simple liquids}.
\newblock Elsevier, 1990.

\bibitem{lange2006collective}
Oliver~F Lange and Helmut Grubm{\"u}ller.
\newblock Collective {L}angevin dynamics of conformational motions in proteins.
\newblock {\em The Journal of chemical physics}, 124(21):214903, 2006.

\bibitem{RevModPhys.48.571}
J.~Schnakenberg.
\newblock Network theory of microscopic and macroscopic behavior of master
  equation systems.
\newblock {\em Rev. Mod. Phys.}, 48:571--585, Oct 1976.

\bibitem{2312}
S.~R. Groot and P.~Mazur.
\newblock {\em Non-equilibrium thermodynamics}.
\newblock Dover Pub., New York, 1984.

\bibitem{Crisanti2012}
Andrea Crisanti, Andrea Puglisi, and Dario Villamaina.
\newblock {Nonequilibrium and information: The role of cross correlations}.
\newblock {\em Phys. Rev. E}, 85(6):061127, 2012.

\bibitem{rohatgi2011webplotdigitizer}
Ankit Rohatgi.
\newblock 2020, webplotdigitizer.
\newblock Online: https://automeris.io/WebPlotDigitizer/.

\bibitem{kheifets2014observation}
Simon Kheifets, Akarsh Simha, Kevin Melin, Tongcang Li, and Mark~G Raizen.
\newblock Observation of {B}rownian motion in liquids at short times:
  instantaneous velocity and memory loss.
\newblock {\em Science}, 343(6178):1493--1496, 2014.

\bibitem{PhysRevA.2.2005}
Robert Zwanzig and Mordechai Bixon.
\newblock Hydrodynamic theory of the velocity correlation function.
\newblock {\em Phys. Rev. A}, 2:2005--2012, Nov 1970.

\bibitem{fodor2015generalized}
{\'E}tienne Fodor, Denis~S Grebenkov, Paolo Visco, and Fr{\'e}d{\'e}ric van
  Wijland.
\newblock Generalized {L}angevin equation with hydrodynamic backflow:
  Equilibrium properties.
\newblock {\em Physica A: Statistical Mechanics and its Applications},
  422:107--112, 2015.

\bibitem{nelder1965simplex}
John~A Nelder and Roger Mead.
\newblock A simplex method for function minimization.
\newblock {\em The computer journal}, 7(4):308--313, 1965.

\bibitem{kuhfittig1978laplace}
Peter~KF Kuhfittig.
\newblock {\em Introduction to the Laplace transform}, volume~8.
\newblock Plenum Press, 1978.

\bibitem{PhysRevE.72.061107}
Jing-Dong Bao, Peter H\"anggi, and Yi-Zhong Zhuo.
\newblock Non-{M}arkovian {B}rownian dynamics and nonergodicity.
\newblock {\em Phys. Rev. E}, 72:061107, Dec 2005.

\bibitem{sekimoto2010stochastic}
Ken Sekimoto.
\newblock {\em Stochastic energetics}, volume 799.
\newblock Springer, 2010.

\end{thebibliography}

\end{document}